\PassOptionsToPackage{unicode}{hyperref}
\PassOptionsToPackage{hyphens}{url}
\PassOptionsToPackage{dvipsnames,svgnames,x11names}{xcolor}
\documentclass[
  12pt]{article}

\usepackage{amsmath,amssymb}
\usepackage{iftex}
\ifPDFTeX
  \usepackage[T1]{fontenc}
  \usepackage[utf8]{inputenc}
  \usepackage{textcomp} %
\else %
  \usepackage{unicode-math}
  \defaultfontfeatures{Scale=MatchLowercase}
  \defaultfontfeatures[\rmfamily]{Ligatures=TeX,Scale=1}
\fi
\usepackage{lmodern}
\ifPDFTeX\else  
\fi
\IfFileExists{upquote.sty}{\usepackage{upquote}}{}
\IfFileExists{microtype.sty}{%
  \usepackage[]{microtype}
  \UseMicrotypeSet[protrusion]{basicmath} %
}{}
\makeatletter
\@ifundefined{KOMAClassName}{%
  \IfFileExists{parskip.sty}{%
    \usepackage{parskip}
  }{%
    \setlength{\parindent}{0pt}
    \setlength{\parskip}{6pt plus 2pt minus 1pt}}
}{%
  \KOMAoptions{parskip=half}}
\makeatother
\usepackage{xcolor}
\makeatletter
\ifx\paragraph\undefined\else
  \let\oldparagraph\paragraph
  \renewcommand{\paragraph}{
    \@ifstar
      \xxxParagraphStar
      \xxxParagraphNoStar
  }
  \newcommand{\xxxParagraphStar}[1]{\oldparagraph*{#1}\mbox{}}
  \newcommand{\xxxParagraphNoStar}[1]{\oldparagraph{#1}\mbox{}}
\fi
\ifx\subparagraph\undefined\else
  \let\oldsubparagraph\subparagraph
  \renewcommand{\subparagraph}{
    \@ifstar
      \xxxSubParagraphStar
      \xxxSubParagraphNoStar
  }
  \newcommand{\xxxSubParagraphStar}[1]{\oldsubparagraph*{#1}\mbox{}}
  \newcommand{\xxxSubParagraphNoStar}[1]{\oldsubparagraph{#1}\mbox{}}
\fi
\makeatother

\usepackage{longtable,booktabs,array}
\usepackage{calc} %
\usepackage{etoolbox}
\makeatletter
\patchcmd\longtable{\par}{\if@noskipsec\mbox{}\fi\par}{}{}
\makeatother
\IfFileExists{footnotehyper.sty}{\usepackage{footnotehyper}}{\usepackage{footnote}}
\makesavenoteenv{longtable}
\usepackage{graphicx}
\makeatletter
\def\maxwidth{\ifdim\Gin@nat@width>\linewidth\linewidth\else\Gin@nat@width\fi}
\def\maxheight{\ifdim\Gin@nat@height>\textheight\textheight\else\Gin@nat@height\fi}
\makeatother
\setkeys{Gin}{width=\maxwidth,height=\maxheight,keepaspectratio}
\makeatletter
\def\fps@figure{htbp}
\makeatother

\makeatletter
\@ifpackageloaded{caption}{}{\usepackage{caption}}
\AtBeginDocument{%
\ifdefined\contentsname
  \renewcommand*\contentsname{Table of contents}
\else
  \newcommand\contentsname{Table of contents}
\fi
\ifdefined\listfigurename
  \renewcommand*\listfigurename{List of Figures}
\else
  \newcommand\listfigurename{List of Figures}
\fi
\ifdefined\listtablename
  \renewcommand*\listtablename{List of Tables}
\else
  \newcommand\listtablename{List of Tables}
\fi
\ifdefined\figurename
  \renewcommand*\figurename{Figure}
\else
  \newcommand\figurename{Figure}
\fi
\ifdefined\tablename
  \renewcommand*\tablename{Table}
\else
  \newcommand\tablename{Table}
\fi
}
\@ifpackageloaded{float}{}{\usepackage{float}}
\floatstyle{ruled}
\@ifundefined{c@chapter}{\newfloat{codelisting}{h}{lop}}{\newfloat{codelisting}{h}{lop}[chapter]}
\floatname{codelisting}{Listing}

\makeatother
\makeatletter
\@ifpackageloaded{caption}{}{\usepackage{caption}}
\@ifpackageloaded{subcaption}{}{\usepackage{subcaption}}
\makeatother

\ifLuaTeX
  \usepackage{selnolig}  %
\fi
\usepackage[]{natbib}
\usepackage{bookmark}

\IfFileExists{xurl.sty}{\usepackage{xurl}}{} %
\hypersetup{
  pdftitle={Title},
  pdfauthor={Author 1; Author 2},
  pdfkeywords={3 to 6 keywords, that do not appear in the title},
  colorlinks=true,
  linkcolor={blue},
  filecolor={Maroon},
  citecolor={Blue},
  urlcolor={Blue},
  pdfcreator={LaTeX via pandoc}}

\usepackage{graphicx}
\usepackage{natbib}
\usepackage{booktabs}
\usepackage{bm}
\usepackage{xurl}
\usepackage{caption}
\usepackage{float}
\usepackage{longtable}
\usepackage{array}
\usepackage{wrapfig}
\usepackage{url}
\usepackage{xcolor}
\usepackage{comment}
\usepackage{microtype}
\usepackage{parskip}
\usepackage{bookmark}
\usepackage{setspace}

\newcommand{\anon}{1}

\begin{document}

\def\spacingset#1{\renewcommand{\baselinestretch}%
{#1}\small\normalsize} \spacingset{1}

\if1\anon
{
  \title{\bf \Large Are Statistical Methods Obsolete in the Era of Deep Learning? A Study of ODE Inverse Problems}
  \author{Skyler Wu\thanks{
    S.W. is partially supported by a Graduate Research Fellowship from the National Science Foundation.}\hspace{.2cm}\\
    Department of Statistics, Stanford University\\
    and \\
    Shihao Yang \thanks{
    S.Y. is partially supported by a National Science Foundation grant DMS-2318883.}\\
    H. Milton Stewart School of Industrial and Systems Engineering,\\ Georgia Institute of Technology\\
    and \\
    S. C. Kou \thanks{
    S.K. is partially supported by an award from Harvard Data Science Initiative.} \\
    Department of Statistics, Harvard University}
  \maketitle
} \fi

\if0\anon
{
  \bigskip
  \bigskip
  \bigskip
  \begin{center}
    {\LARGE\bf Are Statistical Methods Obsolete in the Era of Deep Learning? A Study of ODE Inverse Problems}
\end{center}
  \medskip
} \fi

\bigskip
\begin{abstract}
In the era of AI, neural networks have become increasingly popular for modeling, inference, and prediction, largely due to their potential for universal approximation. With the proliferation of such deep learning models, a question arises: are leaner statistical methods still relevant? To shed insight on this question, we employ the mechanistic nonlinear ordinary differential equation (ODE) inverse problem as a testbed, using the physics-informed neural network (PINN) as a representative of the deep learning paradigm and manifold-constrained Gaussian process inference (MAGI) as a representative of statistically principled methods. Through case studies involving the SEIR model from epidemiology and the Lorenz model from chaotic dynamics, we demonstrate that statistical methods are far from obsolete, especially when working with sparse and noisy observations. On tasks such as parameter inference and trajectory reconstruction, statistically principled methods consistently achieve lower bias and variance, while using far fewer parameters and requiring less hyperparameter tuning. Statistical methods can also decisively outperform deep learning models on out-of-sample future prediction, where the absence of relevant data often leads overparameterized models astray. Additionally, we find that statistically principled approaches are more robust to accumulation of numerical imprecision and can represent the underlying system more faithfully to the true governing ODEs.
\end{abstract}

\noindent%
\vfill

\spacingset{1.8} %

\section{Introduction}

The advancement of deep neural network models in the last fifteen years has profoundly altered the scientific landscape of estimation, prediction and decision making, from the early success of image recognition \citep{krizhevsky2012imagenet, he2016deep}, to the success of self-learning of board games \citep{silver2017mastering}, to machine translation \citep{wu2016google}, to generative AI \citep{ho2020denoising}, and to the success of protein structure prediction \citep{jumper2021highly}, among many other developments. In many of these successes, there are no well-established  mechanistic models to describe the underlying problem --- for example, we do not fully understand how human brains translate from one language to another. As such, it is conceivable that such successes are attributable to deep neural networks' remarkable capabilities for universal function approximation. In contrast, the hand-crafted models that existed before deep neural networks (such as n-gram models \citep{katz1987estimation, brown1992class, bengio2000neural}) were too restricted to offer satisfactory approximation.

How well do deep neural network models work when there are well-established mechanistic models (as in physical sciences, where decades of theoretical and experimental endeavors have yielded highly accurate mechanistic models in many cases) --- in particular, how do the inference and prediction results of deep neural network models compare to more statistical approaches in the presence of reliable mechanistic models --- is an interesting question. The current article focuses on this comparison.

In physical and natural sciences, the mechanistic models are often described by ordinary differential equations (ODEs), which are often referred to as dynamical systems. 
Statistical inference of dynamical systems, also known as the inverse problem \citep{kaipio2006statistical, mueller2012linear}, arises in many scientific studies, where advances in experimental and data collection techniques have allowed researchers to track dynamical systems closer to real-time \citep{xie2015single}. Although ODE models are parametric, their inference has long challenged researchers, as most nonlinear ODEs require numerical solutions, which can be expensive for many practical applications \citep{mueller2012linear}. 

In this article, we compare the neural network approach to the statistical approach on their inference of dynamical systems. Both approaches were proposed to address the numerical challenge of (nonlinear) dynamical systems. Specifically, we compare the performance of the recently developed physics-informed neural network (PINN) (\cite{raissi2017physics, raissi2019physics, lu2021deepxde}), which represents the neural network approach to the ODE inference problem, to that of manifold-constrained Gaussian process inference (MAGI) \citep{yang2021inference}, which represents a principled Bayesian statistical approach to ODE inference. In this comparison, we investigate three different aspects of statistical inference from noisy and sparse observations: (i) how well can the methods recover the model parameters (including quantities of scientific interest defined by the parameters); (ii) how well can the methods recover the system's time-trajectories; and (iii) how well can the methods predict the future state of the system given the observations from a short time window. In the comparison, for a comprehensive understanding we also consider the case of missing system components, where some system components are entirely unobserved---such cases often arise in real experiments due to various (technical or experimental) limitations of data collection.

We consider two dynamical systems in this article: the SEIR model, a widely used model in epidemiology, and the Lorenz model, a model famous for generating chaotic behavior. Together, these two models represent a spectrum of dynamical systems: one with significant scientific implications, and the other characterized by inherent challenges in long-term predictability. These two dynamical systems thus serve as illustrative and meaningful test cases to assess the different inference methods. 

Our study demonstrates that MAGI is generally more robust and reliable than PINN, particularly in sparse and noisy data environments that are common in laboratory sciences. In these environments, researchers often face fewer than 100 observation data points with noise levels ranging from 5\%, 15\% to higher. We found that for the SEIR model, when all system components are observed, the performance of PINN and MAGI is comparable. However, when missing components are present, MAGI clearly outperforms PINN. The contrast becomes even more pronounced with the chaotic Lorenz system, where PINN routinely struggles to handle the complexity of the system, while MAGI continues to consistently produce accurate inference results. Our results indicate that in the era of deep learning, statistically principled methods are still very relevant, especially when working with sparse and noisy data. Our study underscores the importance of proper probabilistic modeling in capturing the underlying dynamics for inference and prediction when one has limited data.

The rest of the article is organized as follows. Section 1.1 briefly discusses the problem of statistical inference of dynamical models. Section 1.2 reviews other approaches. Section 2 provides the architectural details on PINN and MAGI. Section 3 compares PINN and MAGI on the SEIR model. %
Section 4 compares PINN and MAGI on the Lorenz model. %
Finally, Section 5 concludes the article with discussion.

\subsection{Statistical inference of dynamical systems}
In physical and natural sciences, the mechanistic models are often governed by dynamical systems.: 
\begin{equation} \label{eq:ode}
\dot{\bm{x}}(t) = \frac{d\bm{x}(t)}{dt} = f(\bm{x}(t), \bm{\theta}, t), \quad t \in [0, T],
\end{equation}
where the vector \(\bm{x}(t) \) represents the system's state/status at time \( t \), and \( \bm{\theta} \) is the model parameters. 
For example, in neuroscience, ODEs underpin models of neuron signal transmission \citep{Fitzhugh1961, Nagumo1962}. In ecology, predator-prey population dynamics uses ODEs \citep{Lotka1932} to capture interactions between species over time. In systems biology, ODEs map complex gene regulatory networks, enabling insight into cellular responses \citep{Hirata2002}. In epidemiology, ODE models are used to track disease progression and evaluate public health interventions \citep{shaman2012forecasting}. %

Statistical inference of dynamical systems, also known as the inverse problem \citep{kaipio2006statistical, mueller2012linear}, includes estimating the model parameters, inferring/recovering the time-trajectories of the system, and making predictions of the underlying system at future times from noisy observations collected at various time points. %
The difficulty of inferring dynamical system models lies in the computational costs associated with numerically solving the ODEs, %
which has long challenged the applied mathematics community \citep{mueller2012linear}. 

Utilizing the approximating power of neural networks, PINN was recently introduced (\cite{raissi2017physics, raissi2019physics}); it uses a neural network as a function approximator to the solution of the ODE, thus bypassing numerical integration, where the loss function of the neural network incorporates the ODE dynamics. Section 2 describes PINN in detail.

Parallel to the development of PINN, there are also efforts in the statistical community for the inference of dynamical systems. One competitive method is MAGI \citep{yang2021inference}, which assigns a Gaussian process prior on the system components  \(\bm{x}(t)\) and explicitly conditions it to ensure that the derivatives of  \(\bm{x}(t) \) satisfy the ODE specification. This approach enables MAGI to also bypass the need for numerical integration. Section 2 also describes MAGI in detail.

\subsection{Other approaches}

When the function \( f \) in the ODE (Equation \eqref{eq:ode}) is nonlinear, determining \( \bm{x}(t) = \mathbb{X}_{\bm{x}_0, \bm{\theta}}(t)\) given the initial conditions \( \bm{x}_0 \) and the parameter \( \bm{\theta} \) often requires a numerical integration method, such as the fifth-order Runge–Kutta \citep{lapidus1971} $\hat{\mathbb{X}}^{RK5}$. Given access to numerical integration, the inverse problem of estimating $\bm\theta$ from an experimental/observation dataset 
$\{y(\tau_i): i=1,2,\ldots, N\}$, where $y(\tau_i) = X(\tau_i) + \epsilon(\tau_i)$ represents a noisy observation of the system at time $\tau_i$, can be cast as an optimization problem:
\begin{equation}\label{eq:rk5loss}
    (\hat\theta, \hat x_0) = \arg\min_{\theta, x_0} \sum_{i=1}^N \left\| y(\tau_i) - \hat{\mathbb{X}}^{RK5}_{\bm{x}_0, \bm{\theta}}(\tau_i) \right\|^2.
\end{equation}
Executing such an optimization, however, typically requires performing numerical integration repeatedly, which constitutes the key computation bottleneck.

Before the recent development of PINN and MAGI, various attempts have been made to mitigate this computational difficulty. In statistics, an innovative contribution that uses splines as a surrogate for \( \bm{x}(t) \) was introduced in \cite{ramsay2007parameter}, which formulates a generalized profiling procedure that penalizes departure from the ODE dynamics, effectively bypassing repeated numerical integration through a frequentist smoothing spline approach. 
In constrained optimization, surrogate models are also proposed to bypass numerical solvers for ODE inference. Hard constraints for satisfying ODEs are imposed on collocation points --- for example, Gaussian process surrogates are used as convenient interpolation tools with easy computation of the hard constraints \citep{raissi2017machine, spitieris2023bayesian, oates2019bayesian, chen2021solving}. 
The hard constraints, however, are artificially imposed on the Gaussian process rather than a principled Bayesian derivation through conditioning \citep{cockayne2017probabilistic, graepel2003solving, chen2022apik}. Separately, \cite{barber2014gaussian} place a GP prior on $\bm{x}(t)$ and factorize the joint model using a Dirac delta distribution enforcing the ODE; however, this factorization is a pragmatic Bayes approximation that is mathematically incompatible with the actual Bayes law \citep{macdonald2015controversy}.

Other statistical methods \citep{muller2002fitting}, especially the Bayesian model calibration \citep{kennedy2001bayesian}), have also gained popularity \citep{chiachio2021bayesian, stuart2010inverse, stuart2018posterior}. Many of these methods, however, still cannot completely bypass the numerical integration steps \citep{marzouk2007stochastic, lan2016emulation, li2014adaptive, santner2019design}. Those that bypass the numerical integration are mostly two-stage methods \citep{bar1999fitting,muller2004parameter,rai2019gaussian}, where the ODE solution and its derivatives are first reconstructed from observation data using basis function expansion, such as a polynomial basis \citep{franke1998solving} or splines \citep{bar1999fitting}. Then, the differential equation parameters are estimated using regression from the first-step approximation results. Although simple to implement, two-stage methods are not reliable when the data is sparse and noisy \citep{muller2004parameter}: they often require ad hoc follow-up approximations, such as Picard linearization or moment-matching Gaussian distribution for non-linear equations \citep{magnani2022approximate, xun2013parameter, zhou2020inferring}. Rather than bypassing numerical integration, \cite{ghosh2021variational} retain the ODE solver but accelerate Bayesian inference itself by applying variational optimization with the reparameterization trick, achieving significant speed-ups over MCMC.

Two competitive methods for the ODE inference problem proposed in recent years address many of the aforementioned limitations: the neural network-based PINN and a principled Bayesian method MAGI. We describe them in detail in the next section.

\section{PINN and MAGI}

\subsection{PINN: a neural network (NN) approach}
\citet{raissi2019physics} suggests using a neural network (``NN'') $N(t)$ as a function approximator of the system components $\bm{x}(t)$. Their approach incorporate the ODE physics information into the neural network loss function, popularizing the concept of a physics-informed neural network (PINN):
\begin{equation}
    \hat\theta = \arg\min_{\theta, \phi} \underbrace{\frac{1}{M}\sum_{j = 1}^M \left\| \dot{\mathcal{X}}_N(\iota_j) - \dot N_\phi(\iota_j)\right\|^2}_{\text{mechanistic fidelity part}} + \underbrace{\frac{\lambda}{N} \sum_{i=1}^N \left\|y(\tau_i) - N_\phi(\tau_i)) \right\|^2}_{\text{observation loss part}} , \label{eq:pinn}
\end{equation}
where $\dot{\mathcal{X}}_N(t) = f(N_\phi(t), \theta, t)$ approximates $dx/dt$ by plugging in the NN surrogate $N_\phi(t)$ in place of $\bm{x}(t)$, and the gradient $\dot N_\phi$ is computed via automatic differentiation (i.e., backpropagation). The physics loss $\frac{1}{M}\sum_{j = 1}^M \left\| \dot{\mathcal{X}}_N(\iota_j) - \dot N_\phi(\iota_j)\right\|^2$ is essentially a Monte Carlo integration approximation of $\int_t \left\| \dot{\mathcal{X}}_N(t) - \dot N_\phi(t)\right\|^2 dt$ (up to a normalizing constant) using random or pseudo-random collocation/discretization time points $I = \{\iota_j\}_{j=1}^M$. If forecasting is needed, this discretization set $I = \{\iota_j\}_{j=1}^M$ could easily span into the forecasting time horizon. $\phi$ is the over-parameterized NN nuisance parameter (e.g., the weights connecting the neurons). With such a loss function as described in Equation \eqref{eq:pinn}, an NN training process is carried out to obtain the optimized parameters $\theta$. As with any NN-based approach, the architecture, hyper-parameters, and training strategies of PINN require careful tuning and refinement. See \cite{wang2023expert} for a discussion on popular PINN architectures and hyper-parameters, and \citet{rathorechallenges} for a recommended training schedule.

\subsection{MAGI: a statistical Bayesian approach}

The MAGI method, proposed by \citet{yang2021inference}, estimates both the trajectories and parameters of a dynamical system from noisy time-course data without resorting to numerical integration. %
MAGI leverages two central ideas: (i) using a Gaussian process (GP) to represent the system’s trajectories, and (ii) ensuring that these GP representations adhere to the underlying ODE constraints.

As a Bayesian method, MAGI views $\theta$ as a realization of a random variable $\boldsymbol{\Theta}$ from the prior distribution $\pi(\theta)$, and views $x(t)$ as a realization of a GP $X(t)$ with kernel $\mathcal{K}(\cdot, \cdot)$. When $X(t)$ is a GP, its derivative $\dot X(t) = \lim_{h\to 0} \frac{X(t+h) - X(t)}{h}$ is effectively a linear transformation of $X(t)$, which induces a joint Gaussian distribution for $(X(t), \dot X(t))$ as long as the kernel function $\mathcal{K}$ is twice continuously differentiable. The ODE information in Equation \eqref{eq:ode} is captured in MAGI in a Bayesian formulation by conditioning on $W = 0$, where 
\[
W = \sup_{t} |\dot X(t) - {f}(\bm{X}(t), \bm{\Theta}, t)|. \]
Conditioning on $W=0$ ensures that the ODE constraint is enforced. Computationally, $W$ is approximated by discretization at time points $I = \{\iota_j\}_{j=1}^M \subset [0, T]$ 
\begin{equation}
W_I = \max_{t \in I} |\dot X(t) - {f}(\bm{X}(t), \bm{\Theta}, t)| = \max_j |\dot X(\iota_j) - {f}(\bm{X}(\iota_j), \bm{\Theta}, \iota_j)|, \label{eq:WI}
\end{equation}
and the joint posterior distribution of $\{\bm{X}(t)\}_{t=0}^T$ and $\boldsymbol{\Theta}$ is represented by $\bm{X}(t)$ evaluated at $I$. This joint posterior distribution conditioning on the ODE constraint $W_I = 0$ and the noisy observations $y(\tau) \sim \mathcal{N}(x(\tau), \sigma^2)$ can be expressed as
\begin{align}
& \log p\Big(X(I) = x(I), \Theta = \theta ~\Big|~ y(\tau), \, W_I = 0 \Big)  \\
= & \text{ const} +  \log\pi(\theta) + \log p\Big(X(I)\Big) + \log p\Big(y(\tau) \Big| X(I) \Big) + \log p\Big( \dot X(I) = f(X, \theta, I) \Big| X(I) \Big) \\
= & \text{ const} +  \log\pi(\theta) - \frac{1}{2}\Bigg\{ \underbrace{\Big\| x(I) \Big\|_{K^{-1}}^2 }_{\text{GP prior part}}+ \underbrace{\sum_{i=1}^N \frac{1}{\sigma^2}\left\|y(\tau_i) - x(\tau_i)\right\|^2}_{\text{observation error part}} + \underbrace{\Big\| \dot{\mathcal{X}}_G(I) - \dot{\mathcal{G}}(I) \Big\|_{C^{-1}}^2}_{\text{mechanistic fidelity part}} \Bigg\},\label{eq:magi}
\end{align}
where $K = \mathcal{K}(I, I)$ is the kernel matrix at $I$, i.e., the kernel function $\mathcal{K}(\cdot, \cdot)$ evaluated at the discretization time points $I$; $\dot{\mathcal{X}}_G(I) := f(x(I), \theta, I)$ is the right hand side of the ODE evaluated at the GP discretized sample path $x(I)$; $\dot{\mathcal{G}}(I) = \mathcal{'K}(I, I) \mathcal{K}(I, I)^{-1} x(I)$ is the conditional mean of the GP gradient, conditioning on the GP; and the norm matrix $C = \mathcal{K''}(I, I) - \mathcal{'K}(I, I) \mathcal{K}(I, I)^{-1} \mathcal{K'}(I, I)$ is the conditional covariance matrix of the GP gradient conditioning on the GP. Here, $\mathcal{'K} = \frac{\partial}{\partial s} \mathcal{K}(s, t)$, $\mathcal{K'} = \frac{\partial}{\partial t} \mathcal{K}(s, t)$, and $\mathcal{K''}= \frac{\partial^2}{\partial s\partial t} \mathcal{K}(s, t)$; $\|v\|_A^2 = v^\intercal Av$ denotes the quadratic form. %
With Equation \eqref{eq:magi}, MAGI employs Monte Carlo methods to sample $\theta$ and $\bm{x}(\bm{I})$ from the joint posterior. %
If forecasting is needed, the discretization set $I = \{\iota_j\}_{j=1}^M$ will span into the forecasting time horizon, following the recommendation from \citet{wong2023estimating}. Other details of MAGI, including the GP specifications, can be found in \citet{yang2021inference}. When forecasting in a sensitive system (e.g., Lorenz), we propose a \textit{sequential forecasting} routine described in SI Section \ref{sec:magi_configuration} that sequentially expands the discretization set a few timesteps at a time to promote MCMC convergence.

\subsection{Qualitative difference and terminology}

Throughout this article, we use \emph{NN-based} to refer to approaches such as standard PINNs that represent the latent solution $\mathbf{X}(t)$ with a deep neural network and estimate unknowns primarily through loss-based optimization. By \emph{statistically principled} methods, we refer to approaches that specify an explicit probabilistic model for the latent trajectory and observations and carry uncertainty through inference; MAGI is an example, as it begins with a Gaussian-process prior over $\mathbf{X}(t)$ and conditions on both the observations and the differential-equation operator to obtain a posterior distribution over trajectories and parameters. In PINNs, the data-fitting and the ODE residual typically enter as penalty terms in the objective. While squared-error losses can be interpreted as corresponding to Gaussian-noise assumptions, these assumptions are usually implicit rather than part of an explicit generative model. As a result, PINNs are often presented as plug-and-play tools, whereas Bayesian formulations such as MAGI make modeling assumptions and uncertainty explicit, enabling uncertainty quantification and interpretable uncertainty propagation.

In addition, while MAGI is conceptually derived from Bayesian conditioning on the ODE (i.e., enforcing the differential equation as a strict constraint), in practice this conditioning must be implemented on a finite discretization set of time points. Likewise, PINNs enforce the governing equation through a physics-residual term evaluated on a finite set of collocation points. Thus, in the implementations, both methods incorporate ODE fidelity through discretization; the key distinction is how this ODE mechanistic information is balanced against data fit. In MAGI, this balance is handled within an explicit probabilistic model, %
whereas in standard PINNs it is controlled primarily through user-chosen penalty weights (e.g., $\lambda$) and training choices.

\subsection{A note on point estimates and uncertainty quantification}

With its Bayesian foundation and Monte Carlo inference framework, MAGI naturally provides uncertainty quantification for both parameters and trajectories, including posterior credible intervals and predictive uncertainty. In contrast, standard PINNs, as commonly used in practice, primarily return point estimates and point predictions.

This distinction already highlights a practical advantage of MAGI for scientific inference, where uncertainty bands and credible intervals are often necessary for downstream interpretation and decision-making. Since standard PINNs do not natively provide calibrated posterior uncertainty, a direct comparison of uncertainty quantification would be inherently asymmetric (and would require introducing additional, nonstandard uncertainty wrappers for PINN that bring extra modeling choices and calibration procedures). We therefore,
for fair comparisons between PINN and MAGI, %
only compare point estimates of the parameters $\boldsymbol{\theta}$ and trajectories $\mathbf{X}(t)$, taking the posterior means from MAGI's Monte Carlo samples. For PINN, following the general practice, we use the neural network's forward-pass output as its $\mathbf{X}(t)$ estimate, and extract out the corresponding trainable parameters in the neural network architecture when estimating $\boldsymbol{\theta}$. 

\section{The SEIR model}
\label{sec:SEIR}

The Susceptible–Exposed–Infectious–Recovered (SEIR) model is one of the most widely used models in epidemiology. Its popularity stems from its ability to capture key stages of infection dynamics, from early exposure to eventual recovery. In the model, a population is divided into four compartments: Susceptible (S), individuals who can potentially contract the disease but have not yet been infected; Exposed (E), those who have been infected but are not yet infectious (i.e., in the latent stage); Infectious (I), individuals who are capable of transmitting the disease to susceptible hosts; and Recovered (R), those who have recovered or otherwise been removed from the pool of infectious individuals (e.g., by gaining immunity, isolation, or death).

The SEIR model describes infection progression over time and quantifies disease spread in a population by tracking the flow of individuals through these stages ($S$, $E$, $I$ and $R$):%
\begin{equation}
    \frac{dS}{dt} =-\beta \frac{IS}{N}, \quad
    \frac{dE}{dt} =\beta\frac{IS}{N} - \sigma E, \quad
    \frac{dI}{dt} =\sigma E - \gamma I, \quad
    \frac{dR}{dt} =\gamma I , \label{eq:seir_ode}
\end{equation}
where $N \equiv S + E + I + R$ is the total population. %
Three parameters govern the dynamics: the rate of personal contact ($\beta$); the rate of transferring from exposure to infectious ($\sigma$); and the rate of leaving the infectious state ($\gamma$). 
Despite its relatively simple structure, the SEIR model captures the essential features of many real-world outbreaks and serves as a foundation for more sophisticated or specialized epidemiological models \citep{li1995global, kroger2020analytical}.

We compare PINN and MAGI on their performance in inferring the parameters and trajectories of the SEIR model in this section. In the comparison, we generate 100 simulated datasets from the SEIR model, representing the beginning of the transmission of an infectious disease, and then apply PINN and MAGI to each dataset to infer the key quantities of interest, to recover the system trajectories, and to predict the system at future times. Figure \ref{fig:seir_pinn_full_subset} shows one such dataset (the grey dotted points).

One key quantity of interest is the basic reproduction number \( R_0 \), %
defined as $R_0 = \beta/\gamma$ \citep{boonpatcharanon2022estimating}. $R_0$ reflects the expected number of secondary infections caused by a single infected individual in a fully susceptible population; it also relates to the (exponential) rate of decrease of the susceptible population: $S(t) = S(0) \exp[-R_0 (R(t) \,-\,R(0))/N]$. The value of \(R_0\) serves as a fundamental metric to characterize the transmissibility and provides critical thresholds for public health measures. If \( R_0 > 1 \), the disease can invade and persist within a population, potentially leading to large-scale outbreaks or epidemics. If \( R_0 < 1 \), the pathogen’s chain of transmission will be broken, leading to the eventual decline of the disease. %

Two other key quantities of interest are the peak timing and peak intensity of the infectious population, i.e., the time when $I(t)$ will reach its maximum and the corresponding maximum value of $I(t)$. Predicting when the infection level will reach its maximum and how severe that maximum will be allows public health agencies and policymakers to strategically allocate limited resources (such as hospital beds, medical staff, diagnostic tests, vaccine doses, etc.) in anticipation of disease surges. As the datasets that we analyze are from the beginning phase of the disease, where $I(t)$ is still rising, inferring the peak timing and peak intensity requires predicting the system's evolution into future times. 

When comparing the methods, we examine the following inferential tasks:
\begin{itemize}
    \item The accuracy of estimating the model parameters $(\beta, \gamma, \sigma)$, measured by the absolute estimation error (e.g., $|\hat{\beta}-\beta|$) across the 100 independent datasets.
    \item The accuracy of estimating $R_0$, the peak timing, and the peak intensity of $I(t)$, measured by the absolute estimation error (e.g., $|\hat{R}_0-R_0|$) across the datasets.
    \item The accuracy of reconstructing the in-sample system trajectories, measured by root mean squared error (RMSE) for each system component at the observation time points, e.g., 
    \begin{equation}
    \text{RMSE}(E)=\left(\frac{1}{|I_{\text{obs}}|}\sum_{t \in I_{\text{obs}}}\left(\hat E(t) -  E(t) \right)^2 \right)^{1/2}, \label{eq:in-sample RMSE}
    \end{equation}
    where $I_{\text{obs}}=\{\tau_i\}_{i=1}^N$ are the observation time points.
    \item The accuracy of predicting out to future times, measured by RMSE for each system component at future unobserved time points, e.g., 
    \begin{equation}
    \text{RMSE}_{\text{pred}}(E) =  \left(\frac{1}{|I_{\text{future}}|}\sum_{t \in I_{\text{future}}} \left( \hat{E}(t) - E(t) \right)^2 \right)^{1/2}, \label{eq:pred RMSE}
    \end{equation}
    where $I_{\text{future}}$ is the collection of future time points, where we evaluate the prediction.
\end{itemize}

In our study, we consider two distinct scenarios. The first assumes that all components of the SEIR model are directly observed (albeit with noise) at various time points, including the latent \( E \) (exposed) state. We term this scenario as the ``fully observed'' case. Although in real-world applications the $E$ component is rarely observed, it serves as an inference benchmark. The second scenario, which we term the ``missing-component'' case, is more realistic, where the \( E \) component is entirely unobserved. Estimation and prediction of the SEIR system in the missing-component case thus can only use (noisy) data from the accessible \( I \) (infected) and \( R \) (recovered) components.

\subsection{Experimental setup} \label{sec:seir_setup}

Setting the total population size as \( N = 100\% \) and using the relationship $S \equiv 100\% - E -I -R$, we can take an equivalent formulation of the SEIR model using three components $E$, $I$, and $R$. Furthermore, for improved numerical stability and because population proportions are nonnegative, for both PINN and MAGI, we work with an equivalent representation of the SEIR system through three log-transformed components, $\log E$, $\log I$, and $\log R$.

The true parameters of the system are set at \(\beta = 2.0\), \(\gamma = 0.2\), and \(\sigma = 0.6\); these values emulate those of highly contagious infectious diseases (such as measles) \citep{becker1998estimating, paterson2013historical, guerra2017basic, masters2023measles}.
In each simulated dataset, the observation window is from \( t = 0 \) to \( t = 6 \), and a total of 41 equally spaced observations are collected within this time interval. See Figure \ref{fig:seir_pinn_full_subset} for an illustration. Observations are simulated with multiplicative log-normal noise at 15\%, reflecting realistic measurement variability in epidemiological data. The fully observed case has observations on $E$, $I$ and $R$, whereas the missing-component case only has observations on $I$ and $R$.

For both PINN and MAGI, we use a common discretization set of $161$ evenly-spaced time steps in the observation time period of $t \in [0, 6]$ to evaluate the physics-based loss component (see Equation \eqref{eq:pinn}) and $W_I$ (see Equation \eqref{eq:WI}), respectively. For forecasting, an additional $160$ discretization time steps are used for the out-of-sample period of $t \in (6, 12]$. Python codes and the simulated datasets are deposited in an GitHub repository\footnote{Code for reproducing all results can be found at \href{https://github.com/skbwu/stat-vs-dl}{https://github.com/skbwu/stat-vs-dl}.}.

\subsubsection{PINN implementation details}
\label{sec:pinn_implementation}

In PINN, the hyperparameter $\lambda$ governs the balance between the $L_2$ and the physics losses (see Equation \eqref{eq:pinn}). Rather than relying on a single ``out-of-the-box'' PINN setting, we use established recommendations as baselines and systematically sweep the main tunable choices to assess transferability and report best-case performance. We will show and analyze results for $5$ different values of $\lambda$ at $\lambda = 0.1, 1, 10, 100,$ and $1000$.
We also consider two PINN implementations: one based on a practitioner codebase in the PINN literature \citep{van2022physics}, and one using the widely-adopted \texttt{DeepXDE} package \citep{lu2021deepxde}.  Within the \texttt{DeepXDE} framework, we adopt the recommended architecture following the \texttt{DeepXDE} documentations \citep{lu2021deepxde} and examples \citep{wang2022respecting}. In addition to the $5$ values of $\lambda$, we vary 
(i) $2$ network layer-size configurations 
($3$ hidden layers of $40$ or $512$ units); 
(ii) $2$ learning rate scheduler choices 
(constant or exponential decay); and 
(iii) whether to fine-tune with second-order L-BFGS 
after first-order Adam training \citep{rathorechallenges}. Additional PINN details are in SI Section~\ref{sec:pinn_configuration}. For each of the resulting $5 \times 2 \times 2 \times 2 = 40$ configurations, we run $5$ seeds and select the best-performing configuration per $\lambda$ value based on inverse problem accuracy for in-sample experiments, or forecast accuracy for forecasting problems. This post-hoc selection gives PINN an explicit advantage: if MAGI still outperforms the best-case PINN, the conclusion is only strengthened, since it would outperform any other configuration as well. We report these best-case PINN results in the main text and defer comprehensive comparisons across other configurations to SI.

\subsubsection{MAGI implementation details}
\label{sec:magi_implementation}

The MAGI setup follows the recommended procedure of 
\citet{yang2021inference} and is also detailed in SI Section 
\ref{sec:magi_configuration}. The Gaussian process
hyperparameters are estimated by
maximizing the marginal likelihood of the GP smoothing of the observed
data. This automatic procedure is used in all experiments except the 
SEIR missing-component case, where the entirely unobserved $E$ 
component lacks data for marginal likelihood estimation; in that 
setting, the Mat\'{e}rn kernel hyperparameters $(\phi_1, \phi_2)$ for 
$E$ are manually specified, guided by the interpolation and 
optimization routine described in SI Section 
\ref{sec:magi_configuration}. We note that across our PINN 
configuration grid, the missing-component setting is similarly 
challenging: neither method produces reliable results under purely 
automatic tuning.

For the fully observed setting, we perform (a) $1,000$ burn-in and $1,000$ sampling steps for in-sample inference, followed by (b) $1,000$ burn-in and $1,000$ sampling steps for forecasting. In the missing component setting, we perform (a) $10,000$ burn-in and $5,000$ sampling steps for in-sample inference, followed by (b) $5,000$ burn-in and $100,000$ sampling steps for forecasting. These additional steps encourage more thorough exploration of the less-identifiable parameter space.

For numerical stability, we add a small diagonal nugget (jitter) term $10^{-6}$ to GP covariance matrices to ensure numerical stability and invertibility in Cholesky-based computations, as is standard in GP toolkits (e.g., \texttt{GPyTorch}). This nugget is fixed for numerical stability and is not treated as a tunable hyperparameter.

\subsection{Results from the fully observed case}

\begin{figure}[ht!]
    \centering
    \includegraphics[width=1.0\linewidth]{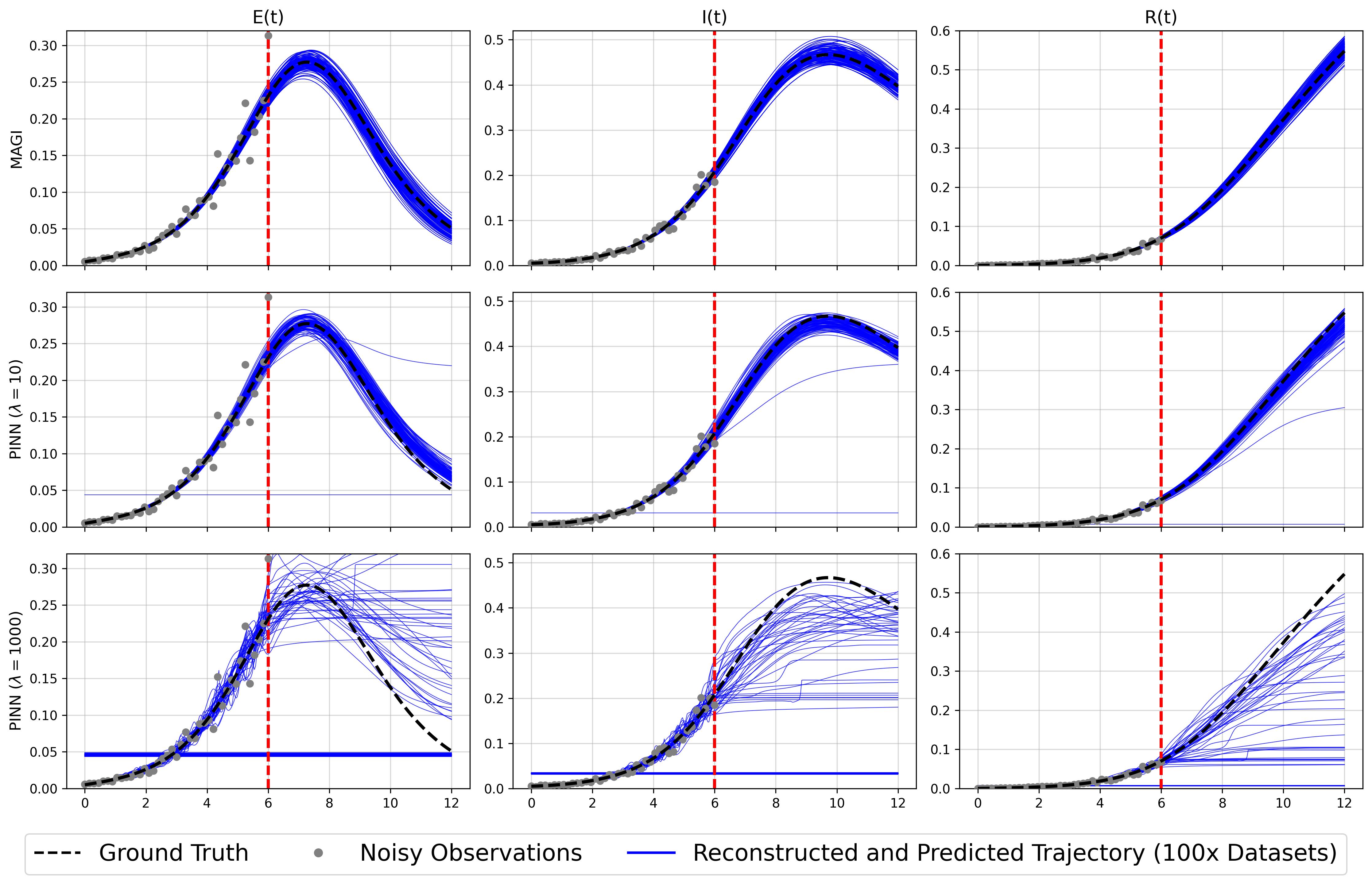}
    \caption{Trajectory reconstruction and prediction by best-case PINN and MAGI for the SEIR model in the fully observed case. The dots show 
    one sample dataset (out of 100). The dashed black lines give the true curves, which are to be identified. Each solid blue curve is the estimate from one dataset. The red dashed vertical line separates the in-sample observation period from the future forecasting period. Top row: MAGI estimates. Lower two rows: best-case PINN estimates, with $\lambda = 10$ and $\lambda = 1000$, respectively. Each solid blue curve is the estimate from one dataset. $S(t)$ is not plotted because it is completely determined once $E(t), I(t)$ and $R(t)$ are known: $S \equiv 100\% - E -I -R$.}
    \label{fig:seir_pinn_full_subset}
\end{figure}

\begin{figure}[ht!]
    \centering
    \includegraphics[width=1.0\linewidth]{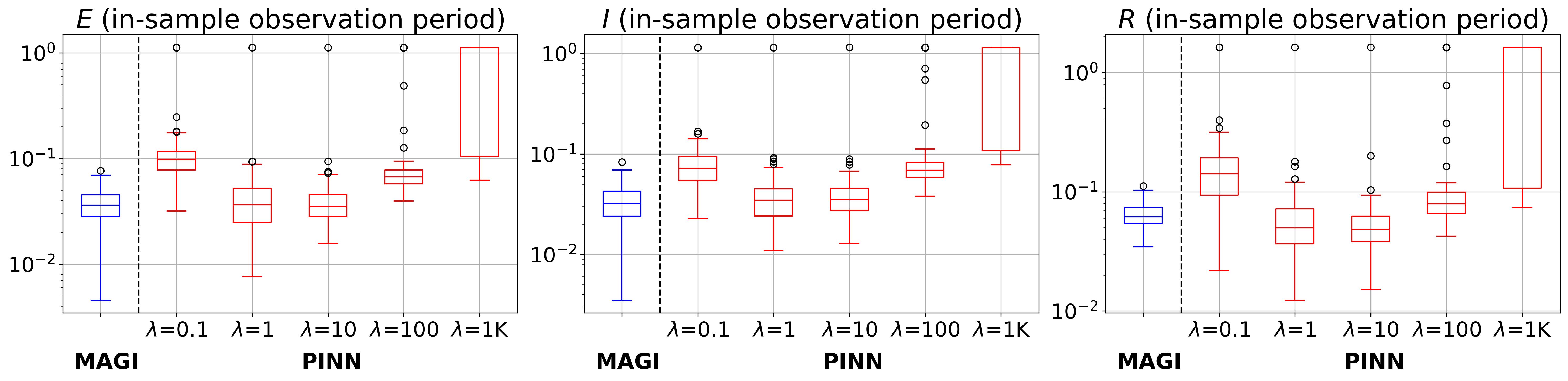} \\
    \includegraphics[width=1.0\linewidth]{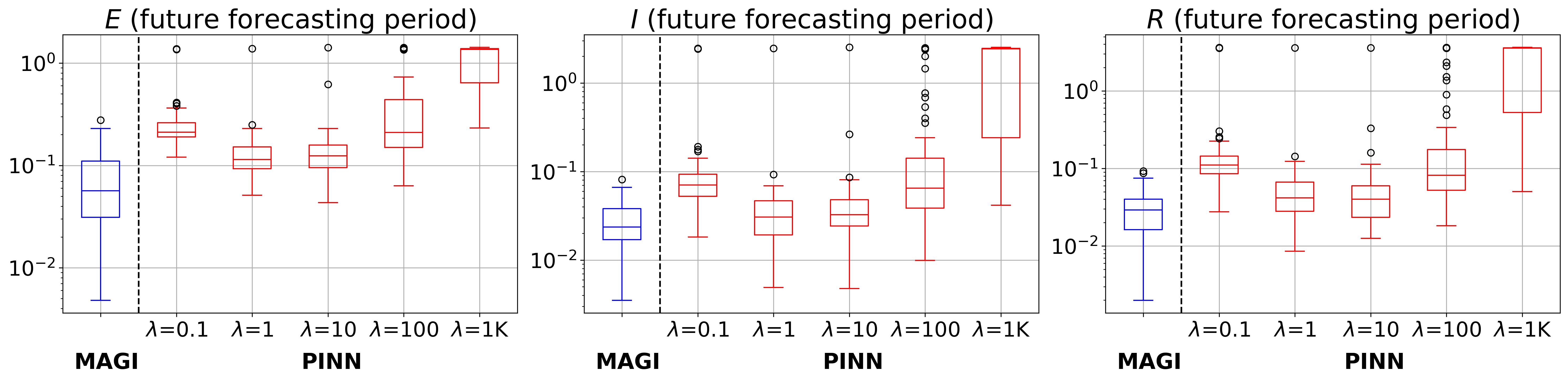}
    \caption{Boxplots showing the RMSE on the logarithm of the SEIR system components across 100 datasets in the fully observed case. Lower value indicates better performance. The y-axis is displayed in the logarithmic scale for better visualization. Top row: in-sample trajectory reconstruction (Equation \eqref{eq:in-sample RMSE}); bottom row: future forecasting (Equation \eqref{eq:pred RMSE}). The three columns correspond to three system components: $E$, $I$, and $R$. In each panel, the leftmost boxplot is for MAGI, and the remaining boxplots are for best-case PINN under different hyperparameters $\lambda$; the dashed vertical line separates MAGI and best-case PINN. Comparison with other PINN configurations can be found in SI Figure \ref{fig:si-full-seir-worse-pinn-boxplot_seir_traj_err_full_logscale}.}
    \label{fig:boxplot_seir_traj_err_full_logscale}
\end{figure}

\begin{figure}[ht!]
    \centering
    \includegraphics[width=1.0\linewidth]{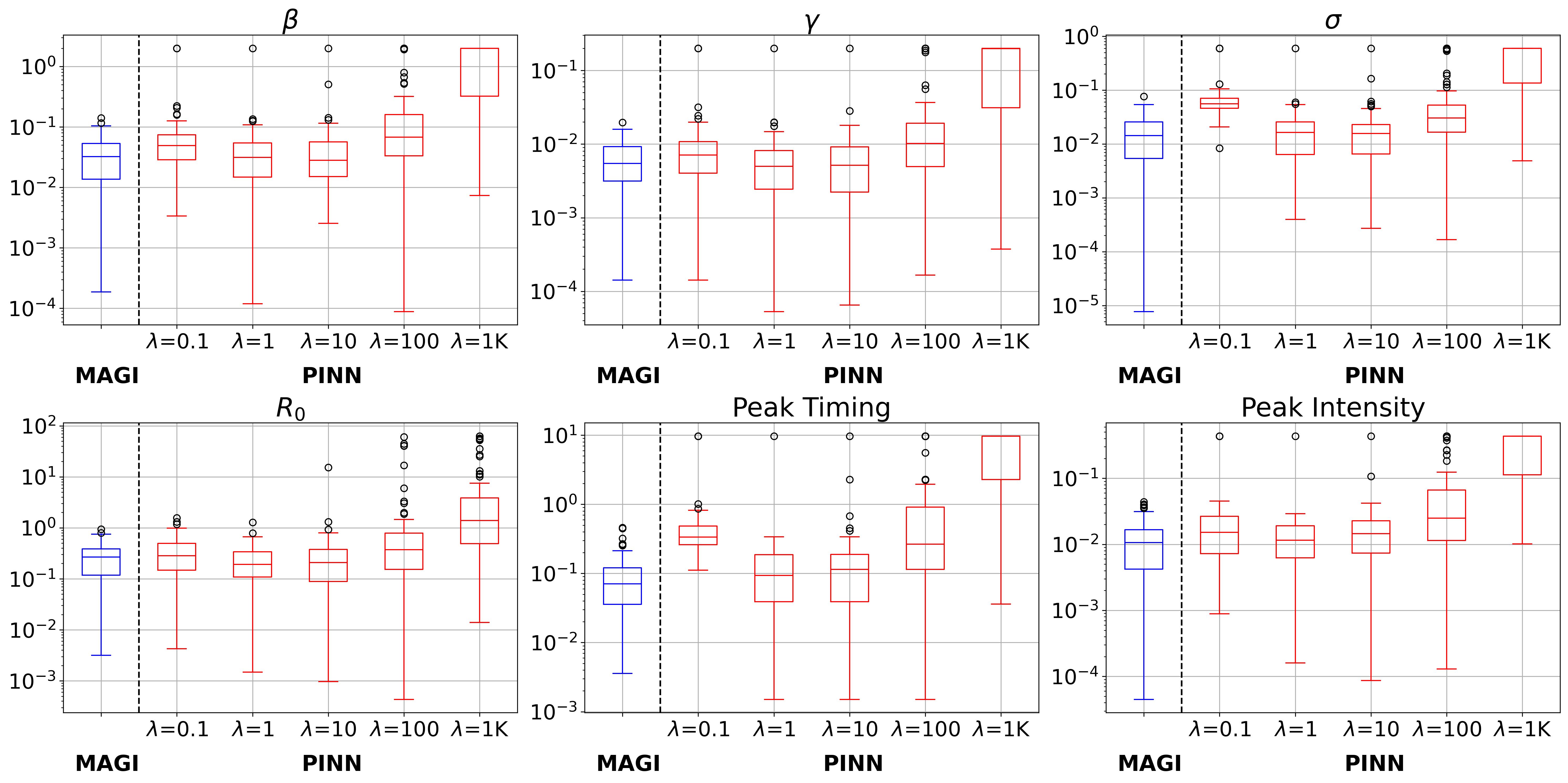}
    \caption{Boxplots showing the absolute errors in parameter estimation for best-case PINN and MAGI across 100 datasets in the fully observed case. Lower value indicates better performance. The y-axis is displayed in the logarithmic scale for better visualization. Top row: The errors for the original parameters $\beta$, $\gamma$, and $\sigma$. Bottom row: The errors for $R_0$, peak timing, and peak intensity -- our three quantities of interest. In each panel, the leftmost blue boxplot is for MAGI, and the remaining red boxplots are for best-case PINN under different $\lambda$ settings; the dashed vertical line separates MAGI and PINN results. Comparison with other PINN configurations can be found in SI Figure \ref{fig:si-full-seir-worse-pinn-boxplot_param_err_full_logscale}.
    }
    \label{fig:boxplot_param_errors_full}
\end{figure}

We start with the fully observed case. We apply PINN and MAGI to each of the 100 simulated datasets. Figure \ref{fig:seir_pinn_full_subset} displays one dataset for illustration. The estimated system trajectories are presented in Figure \ref{fig:seir_pinn_full_subset}, where each solid blue curve corresponds to the result from one dataset. The error metrics for trajectory reconstruction, forecasting and parameter estimation are summarized using boxplots in Figures  \ref{fig:boxplot_seir_traj_err_full_logscale} and \ref{fig:boxplot_param_errors_full}. 

It is evident from Figure~\ref{fig:seir_pinn_full_subset} that the performance of PINN, even under its best-case configuration, is highly sensitive to the choice of the hyperparameter~$\lambda$: in particular, $\lambda = 10$ substantially outperforms $\lambda = 1000$. Section~\ref{sec:addtional_seir_results} in the Supplementary Information further corroborates this observation by examining a broader range of $\lambda$ values as well as other PINN configurations. In the remainder of the paper, we focus on the best-case PINN and, whenever it is clear from context, refer to it simply as PINN. Upon closer examination, even with the best \(\lambda = 10\) setting, PINN occasionally produces outlier trajectories that do not fit the observed data. PINN's tendency to produce outliers is also seen in Figure \ref{fig:boxplot_seir_traj_err_full_logscale}, which contains boxplots of in-sample trajectory reconstruction error (Equation \eqref{eq:in-sample RMSE}) and out-of-sample forecasting error (Equation \eqref{eq:pred RMSE}).
Given that PINN is trained on a large number (60K or 300K) of epochs, such outliers are likely due to the optimizer becoming trapped in local modes. Despite PINN's occasional outliers and greater sensitivity to the choice of the hyperparameter $\lambda$, the majority of well-behaved in-sample trajectory reconstruction instances of PINN and MAGI are roughly comparable, as shown in the top row of Figure \ref{fig:boxplot_seir_traj_err_full_logscale}. However, when we move on to forecasting the system at future times, PINN becomes notably worse than MAGI, as shown in the bottom row of Figure \ref{fig:boxplot_seir_traj_err_full_logscale}. One possible explanation is that, without data to guide PINN at future time points, it struggles to optimize the over-parameterized network to satisfy the ODE. On the contrary, the GP used by MAGI is much more lightweight, making it easier to explore the sampling space and converge towards the correct solution.

The parameter estimation results offer a complementary perspective. Figure \ref{fig:boxplot_param_errors_full} shows the boxplots of the absolute estimating errors for each of the parameters, together with those for the three key quantities of interest. Despite less accurate trajectory forecasting as illustrated in Figure \ref{fig:seir_pinn_full_subset}, PINN is still able to recover key parameters ($\beta$ and $R_0$) competitively to MAGI. This is because, first, the in-sample fittings of PINN and MAGI are comparable, which is more important for parameter estimation. Second, it is well understood from the literature on two-stage approaches for inverse problems that trajectory reconstruction does not have to be perfect in order to recover the ODE parameters \citep{bar1999fitting,muller2004parameter,rai2019gaussian}. The estimated trajectory can be thought of as the response variable, and estimating the ODE parameter can be thought of as carrying out a regression, which could lead to reasonable parameter estimation even though there is noise in the response variable. However, it is also well known that such correspondence becomes less reliable when there are missing components. Indeed, as we will see, the performance of PINN will deteriorate in the missing component case.

\subsection{Results from the missing component case}
\begin{figure}[ht!]
    \centering
    \includegraphics[width=1.0\linewidth]{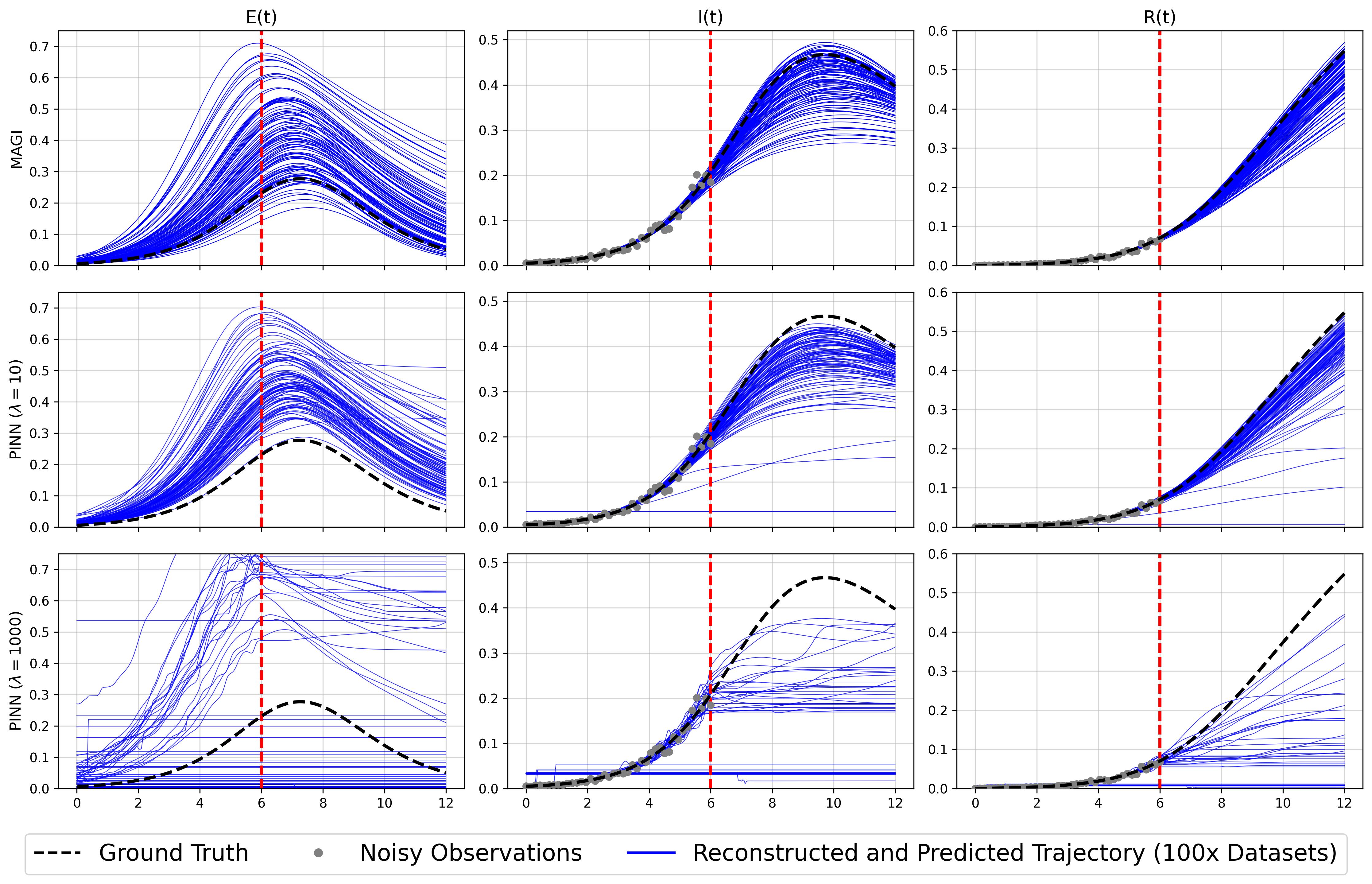}
    \caption{%
    Trajectory reconstruction and prediction by best-case PINN and MAGI for the SEIR model in the missing component case, where the $E$ component is unobserved. The dots show 
    one sample dataset (out of 100). The legend and layout of this figure are identical to Figure \ref{fig:seir_pinn_full_subset}; see the caption there.}
    \label{fig:seir_pinn_partial_subset}
\end{figure}

Observing all the components of the SEIR model is seldom feasible in practice. In this subsection, we consider a more realistic scenario where the exposed component $E$ is not observed, as is typical in infections with an incubation period, and only components $I$ and $R$ are observed. We generate 100 simulated datasets, and visualize one in Figure \ref{fig:seir_pinn_partial_subset}.

When the \(E\) component is missing, a wide range of parameters and latent \(E(t)\) trajectory combinations can produce similar in-sample trajectories for the \(I\) and \(R\) components. This creates a significant challenge for parameter estimation, commonly referred to as the identifiability or ill-posedness issue. As a Bayesian method, MAGI inherently addresses the uncertainty with respect to the missing component through Monte Carlo sampling, which naturally acts as multiple imputations for the unobserved component. 

As shown in Figure \ref{fig:seir_pinn_partial_subset}, the MAGI method fits the in-sample \(I(t)\) and \(R(t)\) accurately while recovering a plausible range of \(E(t)\) from the observed data. The GP prior in MAGI provides stability for the \(E\) component, effectively reducing the variance in its recovery. In contrast, PINN exhibits even greater sensitivity to the hyperparameter \(\lambda\) and produces more noticeable outliers, as shown in Figure \ref{fig:seir_pinn_partial_subset}, and in SI Figure \ref{fig:seir_pinn_partial_plot} for a wider range of $\lambda$.

\begin{figure}[ht!]
    \centering
    \includegraphics[width=1.0\linewidth]{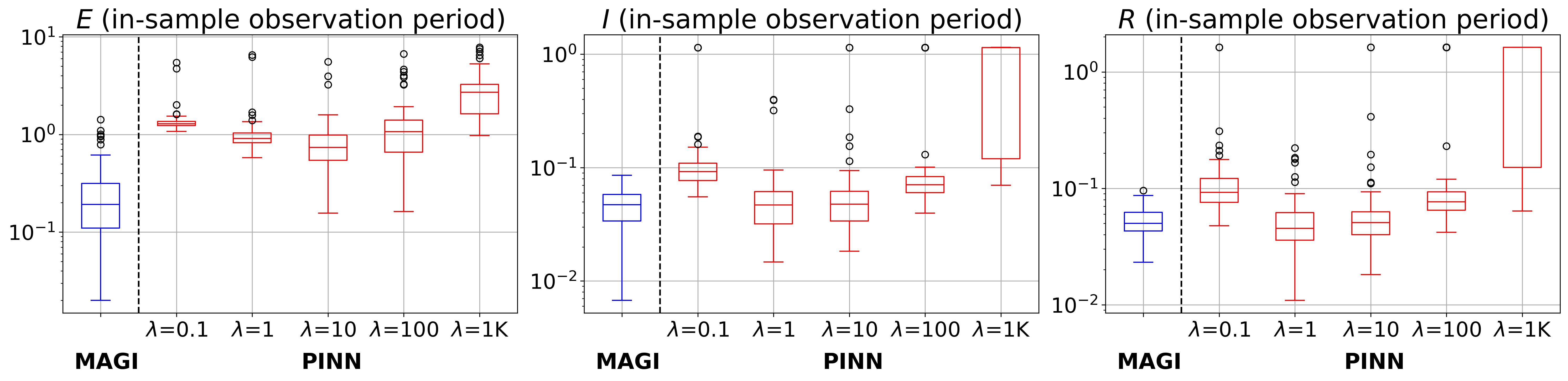}\\
    \includegraphics[width=1.0\linewidth]{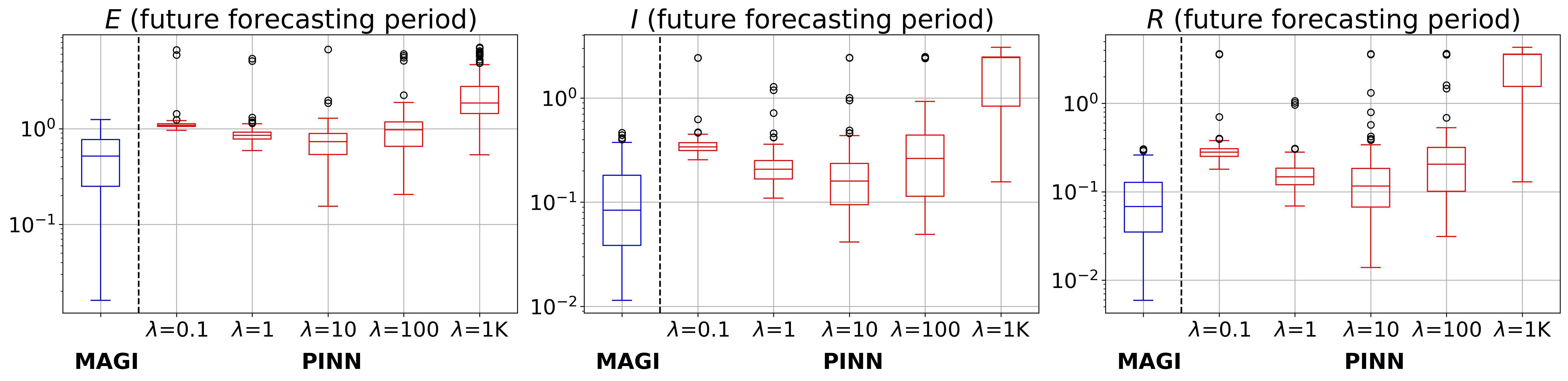}
    \caption{
      Boxplots showing the RMSE on the logarithm of the SEIR system components across 100 datasets in the missing $E$ component case. The legend and layout of this figure are identical to Figure \ref{fig:boxplot_seir_traj_err_full_logscale}; see the caption there. Only best-case PINN configurations are reported here. Results for other PINN configurations are in SI Figure \ref{fig:si-partial-seir-boxplot_seir_traj_err_partial_logscale}.}

\label{fig:boxplot_seir_traj_err_partial_logscale}
\end{figure}

A closer examination of the RMSE boxplots for both the in-sample trajectory reconstruction and future forecasting, presented in Figure \ref{fig:boxplot_seir_traj_err_partial_logscale}, further highlights MAGI’s advantages. MAGI demonstrates a markedly stronger performance compared to PINN, particularly in recovering the latent \(E\) component and forecasting across all components. This observation aligns with findings in the fully observed case, where PINN already struggles with forecasting. The comparable PINN and MAGI RMSE on the observed in-sample \(I\) and \(R\) components is also consistent with the fully observed case. It is when we investigate the recovery of the latent missing $E$ component that we start to see the advantages of the MAGI Bayesian method.

\begin{figure}[ht!]
    \centering
    \includegraphics[width=1.0\linewidth]{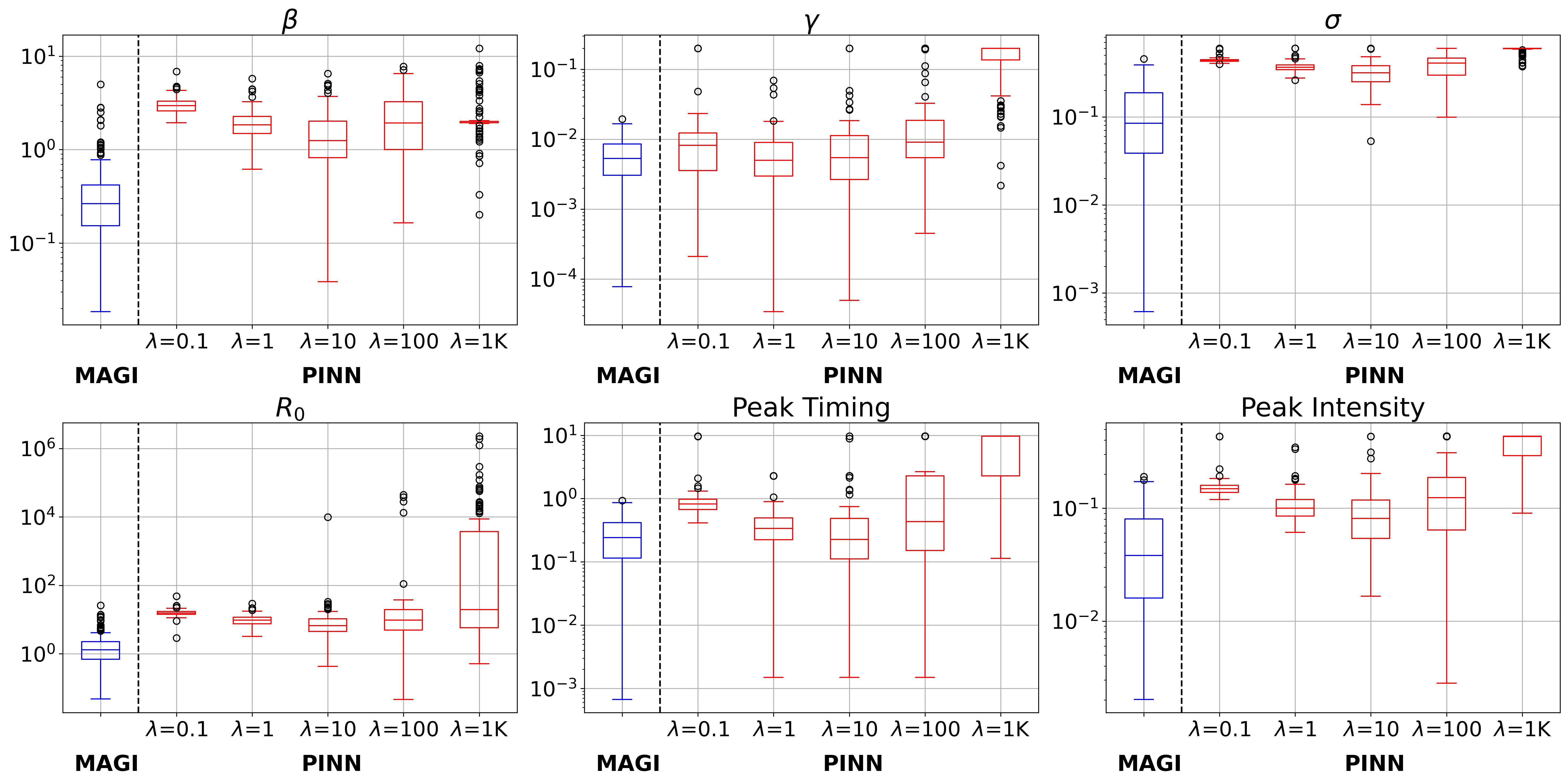}
    \caption{
    Boxplots showing the absolute errors in parameter estimation for PINN and MAGI across 100 datasets in the missing $E$ component case. The legend and layout of this figure are identical to Figure \ref{fig:boxplot_param_errors_full}; see the caption there. Only best-case PINN configurations are reported here. Results for other inferior PINN configurations are in SI Figure \ref{fig:si-partial-seir-boxplot_param_errors_partial}. 
    }
    \label{fig:boxplot_param_errors_partial}
\end{figure}
MAGI's superior performance in recovering the missing $E$ component likely plays a crucial role in its ability to obtain more accurate estimates of the underlying model parameters, as shown in the boxplots of parameter estimation errors in Figure \ref{fig:boxplot_param_errors_partial}. MAGI demonstrates a clear advantage over PINN in estimating the parameters \(\beta\) and \(\sigma\), both of which are directly related to the missing component $E$. In contrast, PINN and MAGI show comparable performance in estimating \(\gamma\), as \(\gamma\) governs the relationship between the observed components \(I\) and \(R\), which do not depend on the missing component. Notably, the estimation of the key quantity of interest, \(R_0\), defined as the ratio \(R_0 = \beta / \gamma\), also shows a significant advantage for MAGI. This observation highlights the necessity of estimating all parameters correctly, not just a subset of them. %
Additionally, MAGI outperforms PINN in predicting peak timing and peak intensity, aligning with the lower RMSE of future trajectory prediction in Figure \ref{fig:boxplot_seir_traj_err_partial_logscale} and the visualization in Figure \ref{fig:seir_pinn_partial_subset}, as estimating the peak timing and peak intensity essentially relies on forecasting of future trajectories.

MAGI's performance results are not without limitations. As a Bayesian approach, MAGI is subject to prior-distribution-induced bias. As shown in Figure \ref{fig:seir_pinn_partial_subset}, MAGI's inferred trajectories of the missing component \(E\) tend to be higher than the ground truth, while the future predictions for \(I\) and \(R\) are slightly lower. This could come from the GP prior, which favors smooth curves around average in-sample values.  %
Interestingly, a similar bias is also observed in the PINN results. Additionally, PINN's occasional outlier predictions resemble a flat line. This phenomenon could be attributable to the NN shrinking toward its initial weights during training, with random initializations prone to outputting flat lines due to variance decay across network layers \citep{glorot2010understanding}. %

\section{The Lorenz model}
\label{sec: Lorenz}

In this section, we compare PINN and MAGI on the Lorenz system, one of the most well-known \textit{chaotic} dynamical systems %
\citep{lorenz1963deterministic, hirsch2012differential, sparrow2012lorenz}. %
Intuitively, a chaotic system is one whose trajectories are extremely sensitive to perturbations in parameters and initial conditions \citep{strogatz2018nonlinear}. %
The Lorenz system has three components $(X, Y, Z)$ governed by the following ODEs, parameterized by $\beta$, $\rho$, and $\sigma$:
$$\frac{dX}{dt} = \sigma (Y - X),\;\; \frac{dY}{dt} = X (\rho - Z) - Y,\;\; \frac{dZ}{dt} = XY - \beta Z.$$ 
The Lorenz system is a good test case due to its chaotic potential, numerical pathologies, and unpredictable behavior. For example, while some Lorenz trajectories will converge to a stationary point (i.e., stable behavior), changing the parameters slightly will yield trajectories that oscillate in the state space in a characteristic butterfly-shaped path for perpetuity without convergence \citep{strogatz2018nonlinear, sparrow2012lorenz}.

\begin{figure}[ht!]
    \centering
    \includegraphics[width=0.5\linewidth]{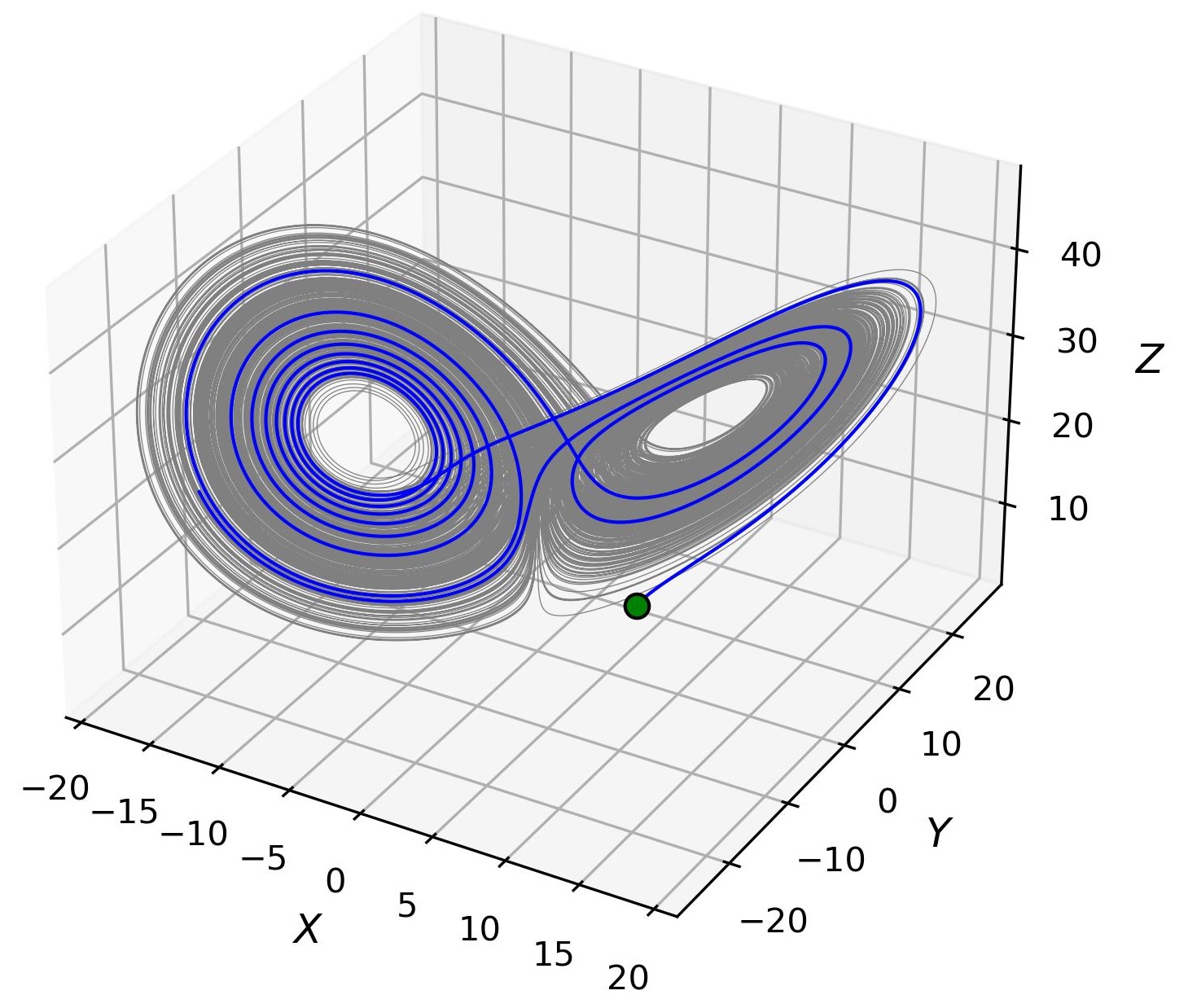} 
    \caption{The Lorenz Chaotic (Butterfly) regime with parameters $\bm{\theta} = (\beta, \rho, \sigma) = (\frac{8}{3}, 28, 10)$ and initial conditions $(X(0), Y(0), Z(0)) = (5, 5, 5)$.}
    \label{fig:chaotic_butterfly}
\end{figure}

We choose a ``Chaotic (Butterfly)'' setting to compare MAGI and PINN. Figure \ref{fig:chaotic_butterfly} depicts the classic chaotic butterfly pattern of the Lorenz system, which, in this case, oscillates in perpetuity. This regime is mathematically chaotic \citep{hirsch2012differential}, with parameters $\bm{\theta} = (\beta, \rho, \sigma) = (\frac{8}{3}, 28, 10)$ and initial conditions $(X(0), Y(0), Z(0)) = (5, 5, 5)$. We generate 100 independent datasets, and then apply PINN and MAGI to each dataset. We consider parameter estimation, in-sample trajectory reconstruction and the prediction of future trajectories in the comparison; the latter two are measured by $\text{RMSE}$ and $\text{RMSE}_{\text{pred}}$ in Equations \eqref{eq:in-sample RMSE} and \eqref{eq:pred RMSE} respectively. Additionally, we also analyzed one stable regime of the Lorenz system and report the results in SI Section \ref{subsec:lorenz_stable_transient_chaos_results}. 

As the Lorenz system could %
be extreme sensitive to perturbations in initial conditions and parameter values, it is mathematically quixotic to predict its trajectories for long periods of time. The purpose of the prediction experiments is thus not to convince the readers to apply either method for forecasting the Lorenz system for long periods, but rather to investigate how \textit{robust} both methods are to the inevitable accrual of floating point imprecision and other numerical errors.

\subsection{Experimental setup}
\label{sec:lorenz-experimental-setup}

We generate 100 independent datasets. The observation window is from $t \in [0, 8]$, and a total of 81 equally spaced observations are generated within this interval in each dataset, which gives $I_{\text{obs}}$. We inject $5\%$ additive Gaussian noise on the observations. %
For both PINN and MAGI, we use a shared discretization set of $I$ as $321$ evenly-spaced time steps in $t \in [0, 8]$ to evaluate the physics-based loss component (see Equation \eqref{eq:pinn}) and $W_I$ (see Equation \eqref{eq:WI}), respectively. 

Separately, for the prediction assessment, we generate another 100 independent datasets; each dataset contains $41$ evenly-spaced observations from $t \in [0, 2]$ with $0.05\%$ Gaussian noise. The specific prediction task is to forecast the system into the future time interval $(2, 5]$ from the observations in $t \in [0,2]$. The prediction is evaluated at $121$ evenly-spaced points on the interval $(2, 5]$, which we denote as $I_{\text{future}}$. We intentionally use a separate low-noise case for prediction. If high noise (e.g., $5\%$) is injected, the chaotic nature of the Lorenz system would render prediction impossible (for any method), making any comparison of computational methods mathematically and numerically meaningless.

We follow very similar PINN and MAGI implementation protocols as in the SEIR experiments (Section~\ref{sec:seir_setup}) to ensure a consistent comparison across systems. For PINN, we enumerate the balancing hyperparameter $\lambda$ in $\lambda = 0.1, 1, 10, 100,$ and $1000$ (Equation~\eqref{eq:pinn}). We evaluate a broad set of PINN configurations by varying the implementation: a practitioner-style codebase used in the PINN literature \citep{van2022physics} versus the \texttt{DeepXDE} package \citep{lu2021deepxde,wang2022respecting}, network size, learning-rate scheduling, and optimizer routine. As in Section~\ref{sec:seir_setup}, we select the best-performing configuration per $\lambda$ value based on inverse-problem accuracy for in-sample experiments, or forecast accuracy for forecasting problems, and only report these best-case PINN results in the main text; comprehensive comparisons across the remaining PINN configurations are deferred to the SI. In the remainder of this Lorenz section, we refer to the best-case PINN simply as PINN whenever it is clear from context.

For MAGI, for all Lorenz experiments, hyperparameters are automatically estimated by marginal likelihood maximization (Section~\ref{sec:magi_implementation}) and are not manually tuned. For in-sample experiments, we perform $3,000$ burn-in and $3,000$ sampling steps. For forecasting, because of the chaotic nature of the Lorenz system, we use the sequential forecasting setup described in SI Section \ref{sec:magi_configuration} with $3,000$ burn-in and $3,000$ sampling iterations in the initial fit and per step. Remaining configuration details are summarized in SI Sections~\ref{sec:pinn_configuration} and~\ref{sec:magi_configuration}.

\subsection{Results from chaotic (butterfly) regime}
\label{sec:lorenz-chaotic-butterfly}

\begin{figure}[ht!]
    \centering
    \includegraphics[width=1.0\linewidth]{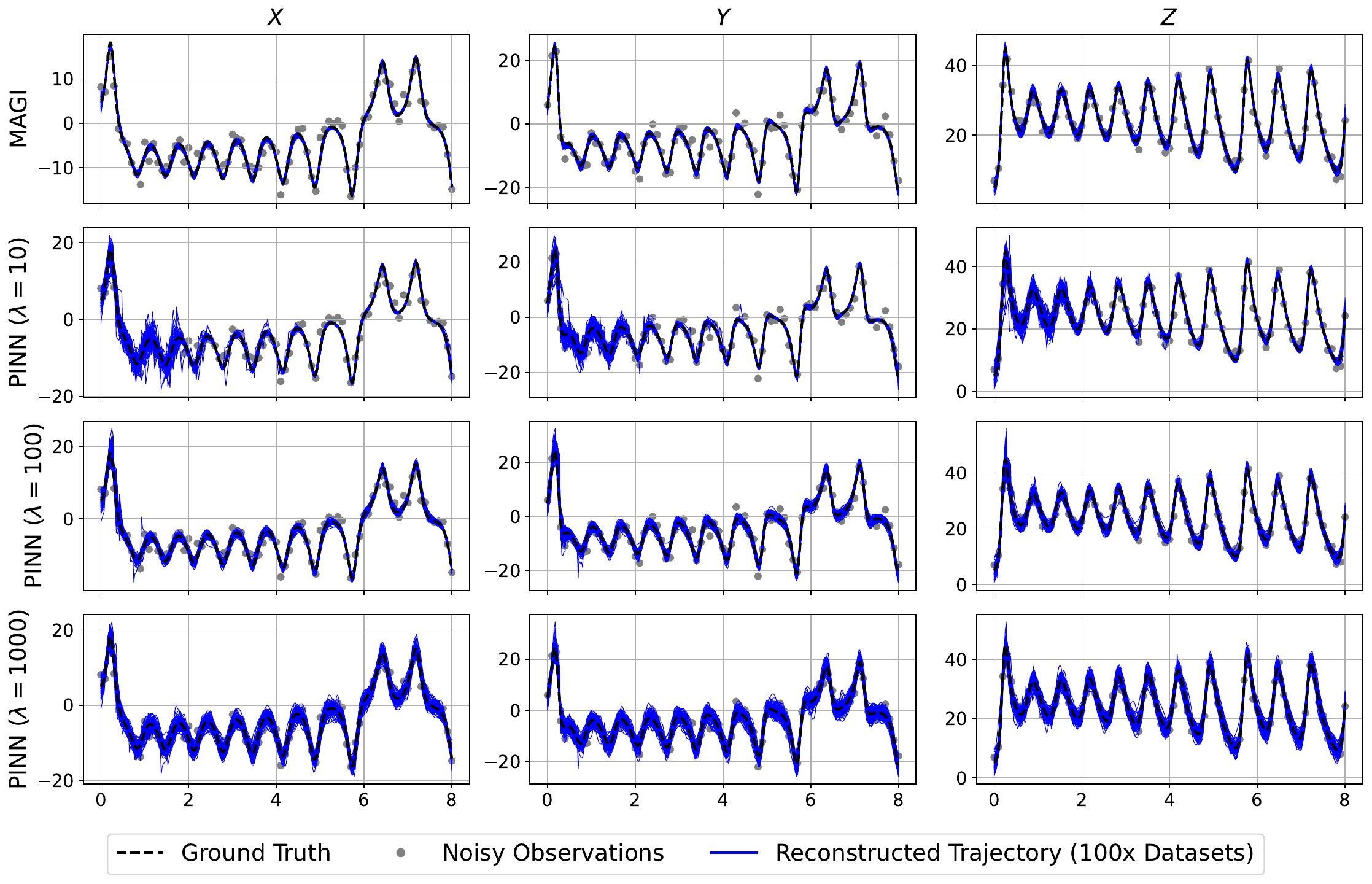} 
    \caption{MAGI and best-case PINN reconstructed trajectories on the Lorenz Chaotic (Butterfly) testbed across $100$ datasets, with one example dataset visualized. Panel layouts are the same as Figure \ref{fig:seir_pinn_full_subset}.}
    \label{fig:chaotic_butterfly_traj_recons_TRUNC}
\end{figure}

\begin{figure}[ht!]
    \centering
    \includegraphics[width=1.0\linewidth]{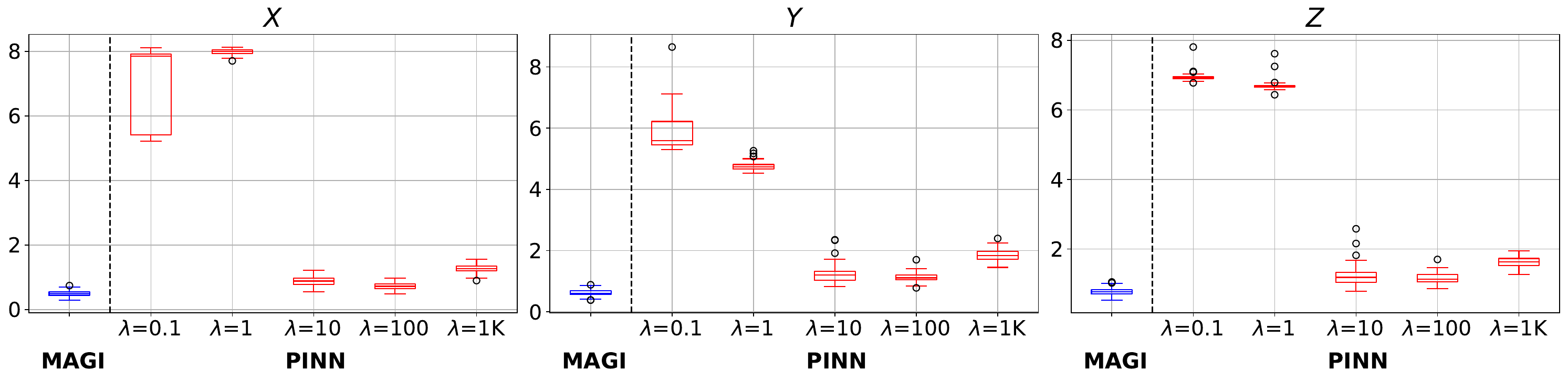} 
    \caption{Boxplots showing trajectory reconstruction RMSEs across $100$ datasets on the Lorenz Chaotic (Butterfly) case. Lower values indicate better performance. The three panels correspond to the three system components $X, Y$, and $Z$. Panel layouts are the same as Figure \ref{fig:boxplot_seir_traj_err_full_logscale}. Only best-case PINN configurations are reported here. Results for other inferior PINN configurations are in SI Figure \ref{fig:si-lorenz-chaotic_butterfly_traj_recons_metrics}.}
    \label{fig:chaotic_butterfly_traj_recons_metrics}
\end{figure}

\begin{figure}[ht!]
    \centering
    \includegraphics[width=1.0\linewidth]{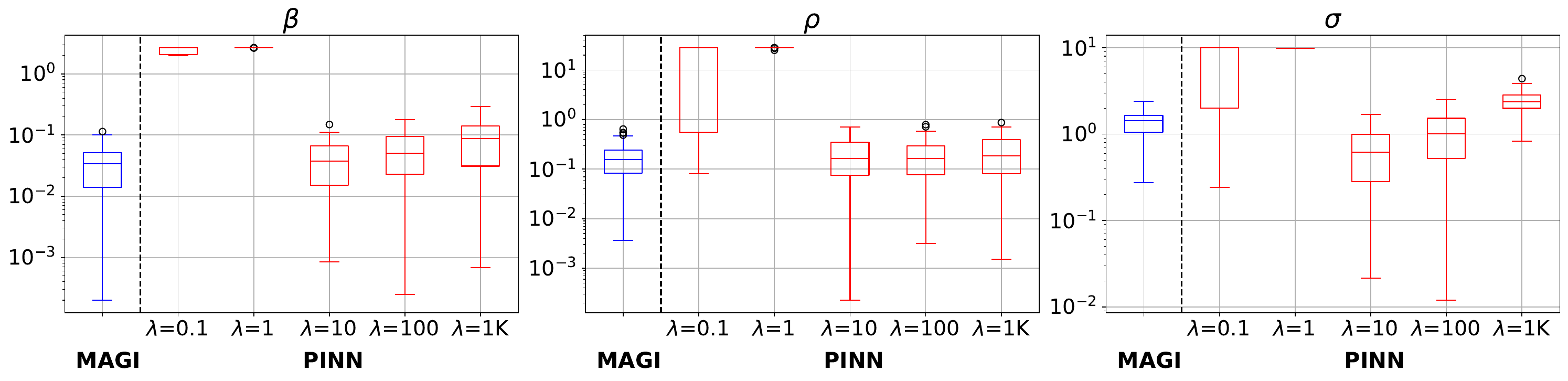} 
    \caption{Boxplots showing parameter inference absolute errors across $100$ datasets on the Lorenz Chaotic (Butterfly) testbed. Lower values indicate better performance. Intra-panel layouts are identical to Figure \ref{fig:chaotic_butterfly_traj_recons_metrics}. Only best-case PINN configurations are reported here. Results for other inferior PINN configurations are in SI Figure \ref{fig:si-lorenz-chaotic_butterfly_param_inf_metrics}.}
    \label{fig:chaotic_butterfly_param_inf_metrics}
\end{figure}

Figure \ref{fig:chaotic_butterfly_traj_recons_TRUNC} displays one dataset out of the 100 for illustration. Figure \ref{fig:chaotic_butterfly_traj_recons_TRUNC} also shows the trajectories inferred by PINN and MAGI from each of the datasets, %
where each blue curve corresponds to the result from one dataset. Errors for in-sample trajectory reconstruction and parameter estimation are summarized using boxplots in Figures \ref{fig:chaotic_butterfly_traj_recons_metrics} and \ref{fig:chaotic_butterfly_param_inf_metrics}.

Visually Figure \ref{fig:chaotic_butterfly_traj_recons_TRUNC} shows that MAGI reconstructs the Lorenz trajectories with both low bias and low trial-to-trial variability: across the $100$ replicated datasets, the reconstructed curves (blue) closely hug the ground-truth trajectory (black dashed) for all three coordinates $(X, Y, Z)$. Moreover, MAGI's uncertainty appears fairly uniform over time: the spread of reconstructions does not noticeably inflate around peaks or troughs. 

In contrast, even the best-performing PINN settings shown above (selected post-hoc:  \(\lambda = 10, 100, 1000\)) exhibit substantially larger variability across trials: reconstructed trajectories look visibly ``fuzzier,'' especially during the initial transient and oscillatory segments. This is likely because these PINNs are overfitting to the noisy data points. Looking across the broader $\lambda$-sweep in Figure \ref{fig:chaotic_butterfly_traj_recons}, smaller $\lambda$ values perform markedly worse: the learned trajectories tend to collapse towards near constant (flat) solutions, effectively attributing most deviations from the flat line to observation noise.

Mathematically, this flat-line output behavior can be explained by the fact that a constant trajectory solution trivially satisfies the governing ODEs of the Lorenz system. Moreover, an over-parameterized neural network with random initialization is prone to producing a flat line output at initialization (especially for small $\lambda$, i.e. underweighting the data-fit term). Gradient-based optimization further exacerbates this issue by naturally inducing shrinkage toward the flat-line initialization, compounding the problem \citep{glorot2010understanding}.

Quantitatively, the RMSE boxplots in Figure \ref{fig:chaotic_butterfly_traj_recons_metrics} corroborate these visual patterns: MAGI achieves lower in-sample reconstruction error than every PINN variant we tested. One plausible explanation is that MAGI's GP priors and principled uncertainty modeling discourage degenerate flat-line solutions and thus favor oscillatory curves that better match the underlying dynamics.

For parameter inference, Figure \ref{fig:chaotic_butterfly_param_inf_metrics} presents boxplots of absolute estimation errors for \((\beta, \rho, \sigma)\). For $\beta$, MAGI tends to have the smallest errors overall compared to all tested PINN settings. For $\rho$, MAGI is broadly comparable to the best post-hoc selected PINN variants (\(\lambda \in \{10, 100, 1000 \}\)), while small-$\lambda$ PINNs perform dramatically worse. Finally, for $\sigma$, certain intermediate-$\lambda$ PINNs (notably, $\lambda = 10$ and $\lambda = 100$) can attain smaller errors than MAGI in this experiment, while overly small or overly large $\lambda$ values degrade performance considerably. Overall, the key takeaway is that PINN performance remains highly $\lambda$-sensitive: underweighting the data-fit term yields large errors consistent with trajectory-collapse behavior, while excessively aggressive data-fitting can impede parameter recovery.

One plausible explanation for PINN's relatively strong performance on $\sigma$ is that while flat-line segments provide little information about the ODE parameters, the periods where PINN successfully captures the oscillatory structure can still reveal meaningful parameter information. As a result, PINN's parameter estimates are not as poor as its trajectory RMSE might suggest. This observation aligns with the SEIR case, demonstrating that PINN, much like a two-stage method, does not require a perfect trajectory estimate to produce reasonable parameter estimates.

Nonetheless, we emphasize that MAGI delivers strong performance without any manual hyperparameter tuning across these Lorenz settings, while PINN requires extensive hyperparameter selection to obtain reasonably comparable results.

\begin{figure}[ht!]
    \centering
    \includegraphics[width=1.0\linewidth]{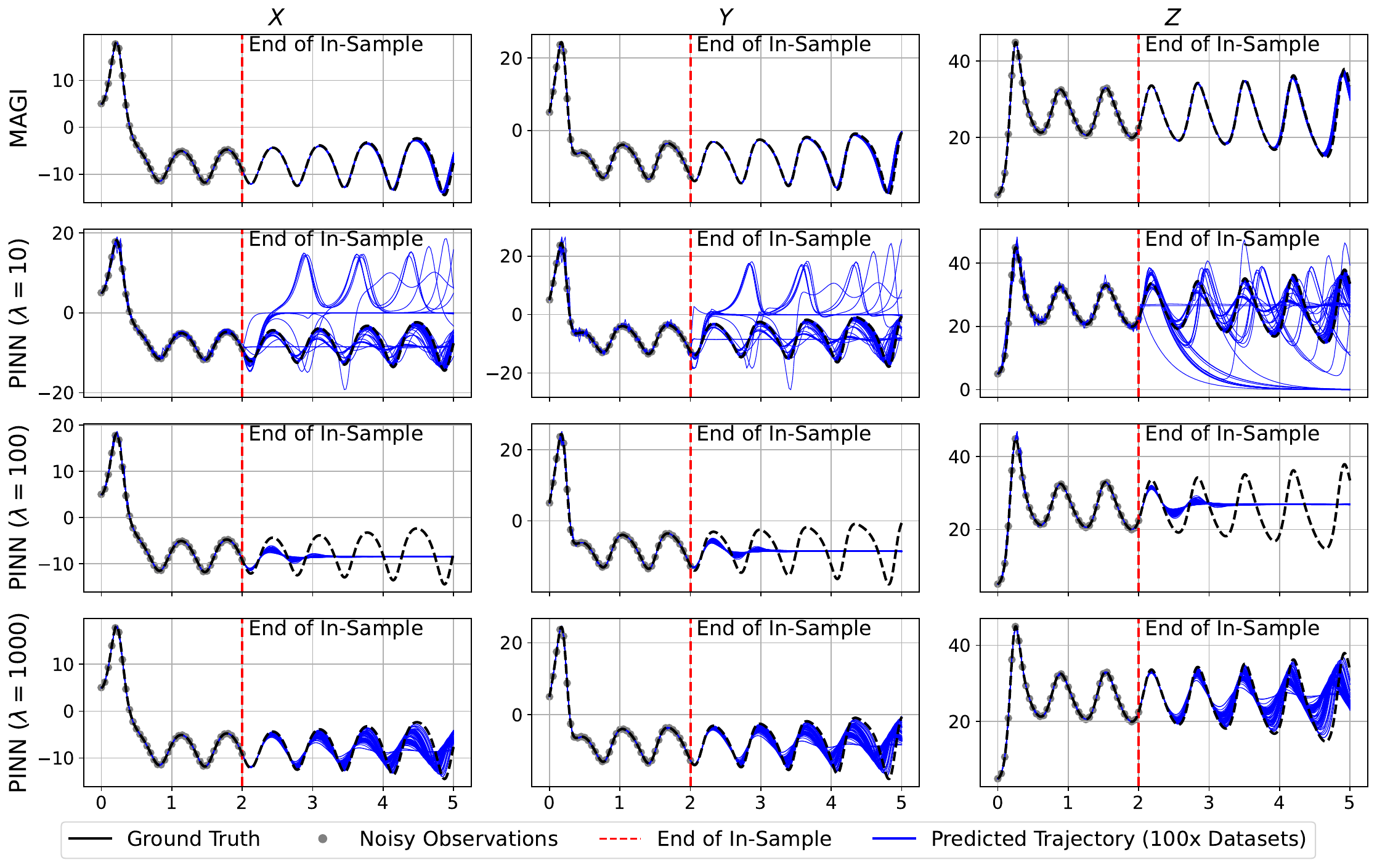} 
    \caption{MAGI and best-case PINN forecasted trajectories on the Chaotic (Butterfly) testbed across $100$ datasets, with one example dataset visualized. The panel layouts are the same as Figure \ref{fig:seir_pinn_full_subset}.}
    \label{fig:chaotic_butterfly_forecasting_TRUNC}
\end{figure}

Finally, Figure \ref{fig:chaotic_butterfly_forecasting_TRUNC} compares the forecasting behavior of MAGI and PINN on the Chaotic (Butterfly) regime. MAGI continues to tightly track the ground-truth trajectory closely through the forecasting window, with relatively little inter-trial dispersion in all three coordinates. In contrast, PINN forecasts are much less stable across datasets and highly sensitive to $\lambda$. For $\lambda = 10$, many PINN forecasts rapidly diverge after the end of the in-sample region and often produce qualitative incorrect forecast trajectories (including collapse towards near-constant behavior in some trials). For $\lambda = 100$, all PINN forecasts appears to collapse towards near-constant behavior. For $\lambda = 1000$, PINN does not collapse as severely, but its forecast spread grows substantially over time, indicative of increased sensitivity/instability in out-of-sample dynamics.

Overall, even in this low-noise setting where PINN can fit the in-sample segment quite well, it struggles to produce reliable forward predictions across $\lambda$ values (see also SI Figure \ref{fig:chaotic_butterfly_forecasting}), consistent with our observation from the SEIR example. In contrast, MAGI can consistently and accurately forecast the future prediction interval up to time $t=5$: beyond this range, the intrinsic sensitivity of chaotic dynamics is expected to impede long-horizon predictability for any numerical method.

It could be argued that the well-known Chaotic (Butterfly) regime of the Lorenz system is notoriously sensitive to parameter variations, making its inverse problem potentially easier. A small change in parameters can lead to significant changes in the system trajectories, implying that observing the trajectories provides strong information about the parameters. To further investigate PINN and MAGI, we thus conducted a second set of experiments under a stable Lorenz regime, where MAGI and PINN performance dynamics are similar to those in the chaotic regime. Detailed analysis of this stable regime can be found in SI Section \ref{subsec:lorenz_stable_transient_chaos_results}.

\section{Discussion and Conclusion}

This study highlights the relevance and potential of statistically principled methods like MAGI in addressing complex systems, even in the era of deep learning. From our analyses on the SEIR and Lorenz systems, it is evident that MAGI routinely outperforms PINN, particularly in more complex scenarios such as those involving missing components and/or chaotic behavior, despite having orders-of-magnitude fewer parameters. Furthermore, as a Bayesian method, MAGI offers uncertainty quantification through credible intervals, a capability that PINN inherently lacks. For instance, on the SEIR model, SI Table \ref{tab:coverage_magi} shows the frequentist coverage of the 95\% credible intervals by MAGI, where the actual coverage is reasonably close to the nominal 95\% level for all system parameters, both in the fully observed case and the missing \(E\) component case.

It is also of interest to investigate how close the reconstructed solutions of the ODEs follow the governing mechanistic equations. To quantify this, we inspect the values of the mechanistic fidelity loss component of PINN in Equation \eqref{eq:pinn} and MAGI's counterpart in Equation \eqref{eq:magi}, using the standard Euclidean norm (instead of over $C^{-1}$) for a meaningful comparison. The results for the Lorenz Chaotic (Butterfly) case are shown in Figure \ref{fig:chaotic_butterfly_mech_fid_metrics}, where we observe that MAGI shows considerably better mechanistic fidelity with respect to the original ODE governing equations than PINN variants with $\lambda > 10$, which are the better-performing PINN configurations for trajectory inference. PINNs with $\lambda = 0.1$ or $1.0$ have slightly better or similar levels of mechanistic fidelity to MAGI, but they sacrifice significant inference performance. Since a flatlined solution trivially satisfies the Lorenz ODEs, mechanistic fidelity is only relevant and useful in combination with strong trajectory inference performance. Nonetheless, MAGI's high mechanistic fidelity---and its apparent robustness to flat-line trajectory collapse---may also contribute to its ability to generalize with future predictions.

Statistical methods like MAGI, derived from a full ground-up probabilistic modeling of the data generating mechanism, can often avoid the need for extensive regularization and ad hoc modifications, as typically required by the training of deep learning methods. For example, MAGI is capable of automatic hyperparameter setting in most cases. On the other hand, PINN can be sensitive to the choice of loss function weighting \citep{dai2024self}. Therefore, we intentionally tested a wide range of configurations for PINN by investigating different choices of (a) the weighting parameter $\lambda$; (b) neural network architecture; (c) learning rate scheduler; and (d) optimizer combinations, and reported the best performing variants per $\lambda$. Despite this effort, PINN oftentimes failed to capture the physics in the Lorenz system. While the sensitivity of PINN to the depth and width of the network typically stabilizes once the network becomes sufficiently large \citep{krishnapriyan2021characterizing}, there are also reported sensitivities to other aspects of the DNN structure, such as the activation function \citep{wang2023learning}, collocation point sampling \citep{wu2023comprehensive}, and other training techniques \citep{cho2024parameterized, wang2024respecting}. Nevertheless, it is evident that PINN requires significantly more careful tuning than MAGI.

The contrast between MAGI and PINN can also be viewed through the lens of what an analyst must specify and what they obtain in return. Bayesian GP-based formulations like MAGI require the analyst to make modeling assumptions explicit (e.g., observation noise, smoothness, and how the differential equation constrains the latent trajectory), but these assumptions are rewarded with coherent uncertainty quantification for parameters, trajectories, and predictions. In contrast, standard PINNs encode the governing equation and data primarily through a penalized optimization objective and are often used as general-purpose tools, but their performance can depend strongly on optimization and tuning choices mentioned above, and they do not natively deliver calibrated posterior uncertainty. More broadly, performance gains arise from exploiting problem structure rather than from universally applicable optimization alone, consistent with the ``No Free Lunch'' perspective that no method can dominate across all problems without leveraging problem-specific assumptions \citep{wolpert1997nofreelunch}.

Now, returning to the question posed in the article title, we argue that statistical methods are still very relevant for inference problems, even in the age of deep-learning and AI. Instead, they provide complementary strengths to neural network approaches, particularly in scenarios with limited amount of data and in scenarios that demand interpretability, robustness, and reliable uncertainty quantification. Our findings in this study are consistent with the long-standing belief that classical statistical or mechanistic models typically outperform neural networks when the model structure is well-specified and accurately reflects the underlying data-generating process. Nevertheless, despite these advantages, there has been a noticeable surge in the popularity of PINN in recent literature, often at the expense of more classical statistical or Bayesian methods such as MAGI, even in cases where the limitations of PINNs are evident. We believe future research that integrates the complementary strengths of statistical methods like MAGI with neural network innovations holds the promise of significantly advancing the field of computational modeling in the era of AI.

\newpage

\section*{Disclosure statement}\label{disclosure-statement}

The authors declare no conflicts of interest.

\section*{Data Availability Statement}\label{data-availability-statement}

Simulated data and detailed implementation code have been made available at \url{https://github.com/skbwu/stat-vs-dl}. (GitHub code repository)

\phantomsection\label{supplementary-material}
\bigskip

\begin{center}

{\large\bf SUPPLEMENTARY MATERIAL}

\end{center}

\begin{description}

\item[Supporting Information:] Configuration of PINN, Configuration of MAGI, Additional table and figures for the discussion section, Additional SEIR result tables and figures, Additional Lorenz result tables and figures. (PDF file)

\item[Code Implementation:] Detailed implementation code at \url{https://github.com/skbwu/stat-vs-dl}. (GitHub code repository)

\end{description}

\clearpage
\begin{spacing}{1.3}
\bibliography{combined}
\end{spacing}

\setcounter{section}{0}

\renewcommand{\thesection}{S\arabic{section}}
\setcounter{figure}{0}
\renewcommand{\thefigure}{S\arabic{figure}}
\setcounter{table}{0}
\renewcommand{\thetable}{S\arabic{table}}
\newpage

\setcounter{page}{1} %

\bigskip
\begin{center}
{\huge\bf Supporting Information}
\end{center}

\section{Configuration of PINN} \label{sec:pinn_configuration}

We explored two versions of PINN, both implemented as fully-connected neural networks: (a) a practitioner-style codebase based on \citet{van2022physics} implemented with TensorFlow 2.17.0 \citep{tensorflow2015-whitepaper}; and (b) a framework using the widely-adopted \texttt{DeepXDE} framework \citet{lu2021deepxde} with a PyTorch 2.10 backend \citep{paszke2019pytorch}.

\subsection{Practitioner-style codebase based on \citet{van2022physics}}
The PINN is implemented as a fully-connected neural network, also referred to as a multi-layer perceptron (MLP). For Lorenz experiments, we use 3 hidden layers with 32 neurons per layer, with batch normalization. For SEIR experiments, we use 5 hidden layers with 32 neurons per layer, with batch normalization. In both cases, the networks are trained using the Adam optimizer with a fixed learning rate of $0.01$ for $60,000$ epochs. The activation function for all layers is the hyperbolic tangent ($\tanh$), chosen for its smoothness and suitability for modeling continuous physical processes. We sweep over 5 values of the $L_2$-vs.-physics-loss governor hyperparameter $\lambda \in \{ 0.1, 1.0, 10.0, 100.0, 1000.0 \}$. ODE parameters for both Lorenz and SEIR are modeled on the log-scale. ODE components for Lorenz are modeled on the original scale, while ODE components for SEIR are modeled on the log-scale for numerical stability.

Our implementation is gratefully borrowed from \cite{van2022physics}'s formulation of PINN for the Lorenz system.

\subsection{DeepXDE-based framework based on \cite{lu2021deepxde}}

The PINN is also implemented as a fully-connected neural network. We sweep over 5 values of the $L_2$-vs.-physics-loss governor hyperparameter $\lambda \in \{ 0.1, 1.0, 10.0, 100.0, 1000.0 \}$. We also sweep over (i) 2 network layer-size configurations (3 hidden layers of 40 or 512 units); (ii) 2 learning rate scheduler choices for initial first-order Adam optimization (constant at $0.001$ or exponentially decaying by $0.9$ every $5,000$ epochs with initial learning rate $0.001$); (iii) whether to fine-tune with second-order L-BFGS after first-order Adam \citep{rathorechallenges}. If a fixed learning rate is used, we run Adam for $60,000$ epochs; if the exponentially-decaying learning rate is used, we run Adam for $300,000$ epochs, following the best practices demonstrated in \cite{lu2021deepxde} and \cite{wang2024respecting}. ODE parameters for both Lorenz and SEIR are modeled on the log-scale. ODE components for Lorenz are modeled on the original scale, while ODE components for SEIR are modeled on the log-scale for numerical stability. We do note that the fully-connected neural network for SEIR outputs with a log-softmax transformation to ensure that $S+E+I+R = 1$ with $S,E,I,R \geq 0$: this is necessary to ensure numerical stability of \texttt{DeepXDE}.

To emphasize, we select the best (i.e., lowest parameter error for in-sample problems or forecasting error for forecasting problems) PINN hyperparameter combination per $\lambda$ over 5 seeds to make the comparison conservative in PINN’s favor and then re-run the selected configurations for a total of 100 seeds.

The full implementation details and codebases for the PINNs are available at the following anonymized GitHub repository: \href{https://anonymous.4open.science/r/tas-revisions-7650}{https://anonymous.4open.science/r/tas-revisions-7650}.

\section{Configuration of MAGI} \label{sec:magi_configuration}

For the implementation of MAGI, we utilize a Gaussian process (GP) kernel based on the Matérn kernel with 2.01 degrees of freedom. This kernel choice balances flexibility and smoothness, making it well-suited for modeling the dynamics of the system. The implementation is built using TensorFlow Probability 0.25.0 (\cite{dillon2017tensorflow}), leveraging its built-in advanced Markov Chain Monte Carlo capabilities.

The hyperparameters of the Gaussian process are estimated through GP smoothing of the observed data, as previously documented in \citet{yang2021inference}. For missing components of the system, the hyperparameters are determined through one of two strategies: (1) setting them to fixed values; or (2) employing a novel interpolation and optimization routine discussed in the following subsection. Please see Section \ref{forecasting_details} for forecasting details.

For Lorenz, ODE components are sampled on the original scale, while for SEIR, they are sampled on the log-scale. ODE parameters for both the Lorenz and SEIR testbeds are sampled on the original scale. Numbers of MCMC burn-in and sampling steps are described in Sections \ref{sec:magi_implementation} and \ref{sec:lorenz-experimental-setup}. For Lorenz (Chaotic Butterfly), Lorenz (Transient Chaos), and SEIR (fully observed components), we use MAGI's automatic hyperparameter tuning routines (no user-side intervention needed). We only manually specify hyperparametres for the SEIR (missing component) setting.

For apples-to-apples comparisons between PINN and MAGI, we will only compare point estimates of the parameters $\boldsymbol{\Theta}$ and trajectories $\mathbf{X}(t)$, i.e., taking the posterior means from MAGI's Monte Carlo samples. For PINN, we use the neural network's forward-pass output as its $\mathbf{X}(t)$ estimate, and extract out the corresponding trainable parameters in the neural network architecture when estimating $\boldsymbol{\Theta}$.

The full implementation details and codebase for MAGI are available at the following anonymized GitHub repository: \href{https://anonymous.4open.science/r/tas-revisions-7650}{https://anonymous.4open.science/r/tas-revisions-7650}

\subsection{Improvements on \cite{yang2021inference}}

Building on the algorithm introduced in \cite{yang2021inference}, we include the following improvements through this new TensorFlow Probability implementation.
\begin{enumerate}
    \item First, we upgraded the base Hamiltonian Monte Carlo (Hamiltonian Monte Carlo \citep{neal2012mcmc}) (HMC) sampler from \cite{yang2021inference} to the No U-Turn Sampler (\cite{hoffman2014no}), itself a variant of HMC, with Dual-Averaging Stepsize Adaptation (\cite{nesterov2009primal}) to enable a completely tuning-free setup, leveraging TensorFlow Probability's collection of advanced Monte Carlo samplers. Our implementation is fully-compatible with GPUs and XLA integration for maximum computational power. For future-proofing, we include a toggleable option for logarithmic annealing of the log-posterior function to encourage additional exploration of the state space, but this is not enabled in any of our experiments.
    \item Second, we implement GP smoothing of the data towards fitting the Matérn kernel hyperparameters using TensorFlow Probability's GaussianProcess class coupled with the Adam optimizer for increased numerical stability.
    \item Third, we add $1 \times 10^{-6}$ perturbation to the diagonal entries of the $\mathcal{K''}$ matrix for increased numerical stability and invertibility.
    \item Fourth, in addition to the Fourier-based Matérn kernel hyperparameter prior introduced in \cite{yang2021inference}, we also provide the option to use a flat prior in cases where the system is not approximately oscillatory. In this study, we enable the Fourier prior on the Lorenz system, but disable it on the SEIR system.
    \item Fifth, we introduce a novel procedure for interpolating missing components' observations and fitting corresponding kernel hyperparameters, all without the need for manual user intervention. First, we use a second-order gradient-matching procedure optimized via Adam to jointly estimate the missing components' trajectory values and the initial values for $\boldsymbol{\Theta}$. Second, we apply GP smoothing on the interpolated missing components' trajectory values to estimate the kernel hyperparameters. This is a departure from the less-stable method used in \cite{yang2021inference} that directly tries to maximize the MAGI log-posterior function, and in the spirit of modern ``differentiable almost everything'' paradigms in machine learning (see \href{https://differentiable.xyz/}{ICML 2024 Workshop on Differentiable Almost Everything}).
    \item Finally, to provide smoother initializations of $\mathbf{\hat{X}}(t)$ for the NUTS sampler and reduce the risk of the sampler becoming stuck in local modes, we introduce a cross-validated cubic-spline-based smoothing mechanism to reduce jaggedness in the original linearly-interpolated initializations of $\mathbf{\hat{X}}(t)$, which also does not require manual user intervention.
\end{enumerate}
The complete codebase for the improved MAGI algorithm is available at the following anonymized GitHub repository: \href{https://anonymous.4open.science/r/tas-revisions-7650}{https://anonymous.4open.science/r/tas-revisions-7650}.

\subsection{Forecasting}
\label{forecasting_details}

Practically, we perform forecasting on MAGI on the Lorenz testbeds via a \textit{sequential forecasting} routine. We begin by obtaining an initial in-sample fit from $t=0$ to $t=2$, setting $I$ as $81$ evenly-spaced timesteps on $t \in [0, 2]$. Then, for our first \textit{sequential forecasting} step, we attempt to forecast from $t=2$ to $t=3$. To accomplish this, we append $I$ with $40$ evenly-spaced time points for $t \in (2, 3]$. We warm-start our initial values for $\boldsymbol{\Theta}$ using the last NUTS sample from our initial in-sample fit. For $\mathbf{\hat{X}}(t)$, for $t \in [0, 2]$, we warm-start our sampler with our last sampled trajectory from our in-sample fit, and warm-start our trajectory for $t \in (2, 3]$ via numerical integration, using our last sample of $\boldsymbol{\Theta}$ and $\mathbf{\hat{X}}(2)$ as our initial conditions. We also refit $\boldsymbol{\phi}$ using the last length-$1$ interval trajectory from the posterior mean obtained from our initial fit. Then, having updated our $I$, sampler initial values for $\boldsymbol{\Theta}$ and $\mathbf{\hat{X}}(t)$, and our kernel hyperparameters $\boldsymbol{\phi}$, we run the NUTS sampling routine again. We repeat this \textit{sequential forecasting} routine until we have forecasted our desired interval length. For the initial fit and each forecasting step, we will use $3000$ burn-in steps followed by $3000$ sampling steps at each stage.

\section{Additional table and figures for discussion}

\begin{table}[H]
\small
\caption{Frequentist coverage of the 95\% credible interval by MAGI under the two SEIR cases: fully observed and missing component.}
\label{tab:coverage_magi}
\centering
\begin{tabular}{l|ccc}
\toprule
parameter & $\beta$  & $\gamma$  & $\sigma$  \\
\midrule
Fully Observed Case & 90\% & 89\% & 90\% \\
Missing $E$ Component Case & 94\% & 91\% & 93\% \\
\bottomrule
\end{tabular}
\end{table}

\begin{comment}
%
\begin{figure}[H]
    \centering
    \includegraphics[width=1.0\linewidth]{figs/lorenz/trajectories/trajs_over_epoch.jpg} 
    \caption{PINN ($\lambda = 1000$) inferred trajectories over epoch on Lorenz (Chaotic Butterfly), with $81$ evenly-spaced observations at $5\%$ level noise from $t=0$ to $t=8$. Dashed, dash-dotted, and solid lines represent inferred trajectories after 12K, 36K, and 60K epochs. The three panels correspond to the three system components: $X$, $Y$, and $Z$.}
    \label{fig:neural_network_slow_trajs}
\end{figure}
\end{comment}

%
\begin{figure}[ht!]
    \centering
    \includegraphics[width=1.0\linewidth]{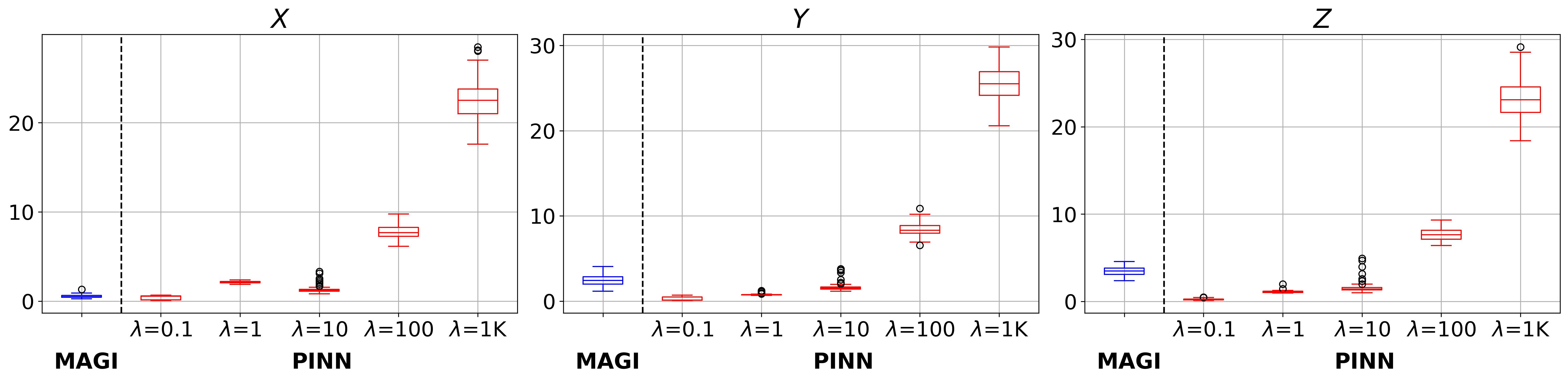} 
    \caption{Chaotic (Butterfly) mechanistic fidelity errors from $100$ datasets, each with $81$ evenly-spaced observations at $5\%$ level noise from $t=0$ to $t=8$. Lower values indicate better mechanistic fidelity. The three panels correspond to the three system components: $X$, $Y$, and $Z$. In each panel, the leftmost boxplot is for MAGI, and the remaining boxplots are for PINN under different hyperparameters $\lambda$; the dashed vertical line separates MAGI and PINN.}
    \label{fig:chaotic_butterfly_mech_fid_metrics}
\end{figure}

\section{Additional SEIR System Results}
\label{sec:addtional_seir_results}

\begin{figure}[H]
    \centering
    \includegraphics[width=1.0\linewidth]{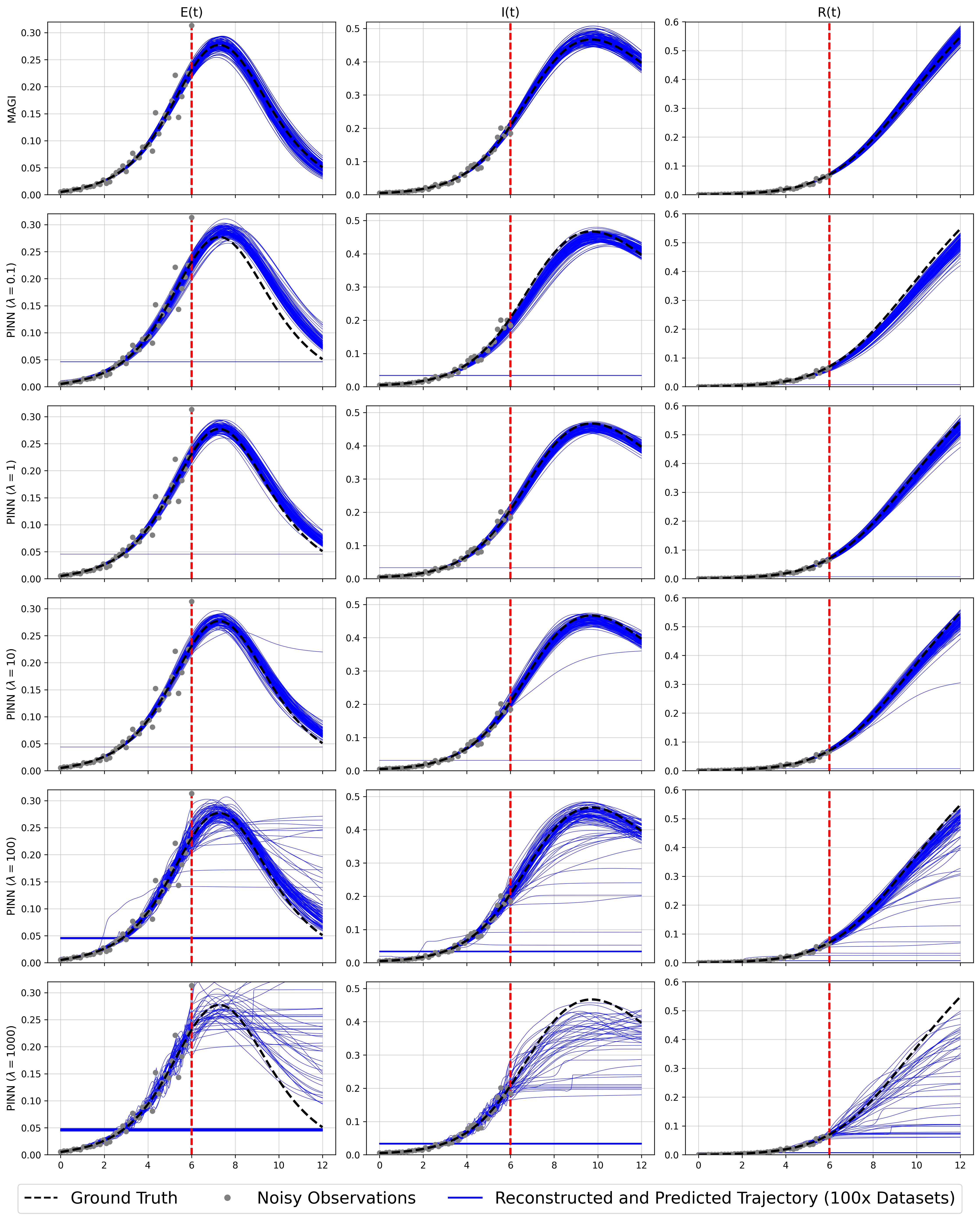}
    \caption{Trajectory reconstruction and prediction by best-case-per-$\lambda$ PINN and MAGI for the SEIR model in the fully observed case; for this SEIR example, the 
    practitioner-style implementation \citep{van2022physics} 
    outperformed the \texttt{DeepXDE} implementation across all 
    $\lambda$ values. The dots show 
    one sample dataset (out of 100). The dashed black lines give the true curves, which are to be identified. The red dashed vertical line separates the in-sample observation period from the future forecasting period. Each solid blue curve is the estimate from one dataset. Top row: MAGI estimates. Lower five rows: best-case-per-$\lambda$ PINN estimates over all five tested $\lambda$ hyperparameter values.}
    \label{fig:seir_pinn_full_plot}
\end{figure}

\clearpage
\begin{figure}[t!]
    \centering
    \includegraphics[width=1.0\linewidth]{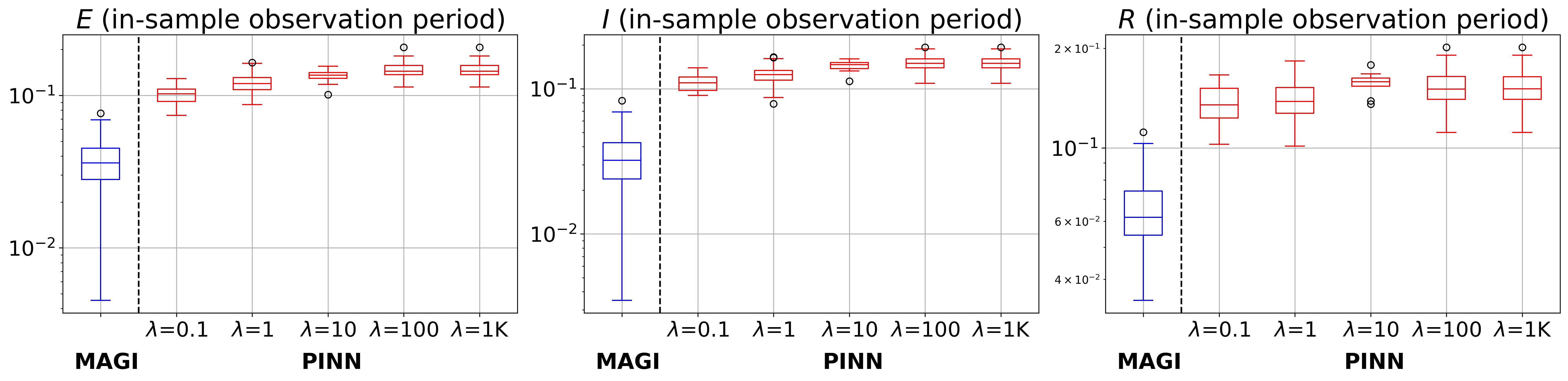} \\
    \includegraphics[width=1.0\linewidth]{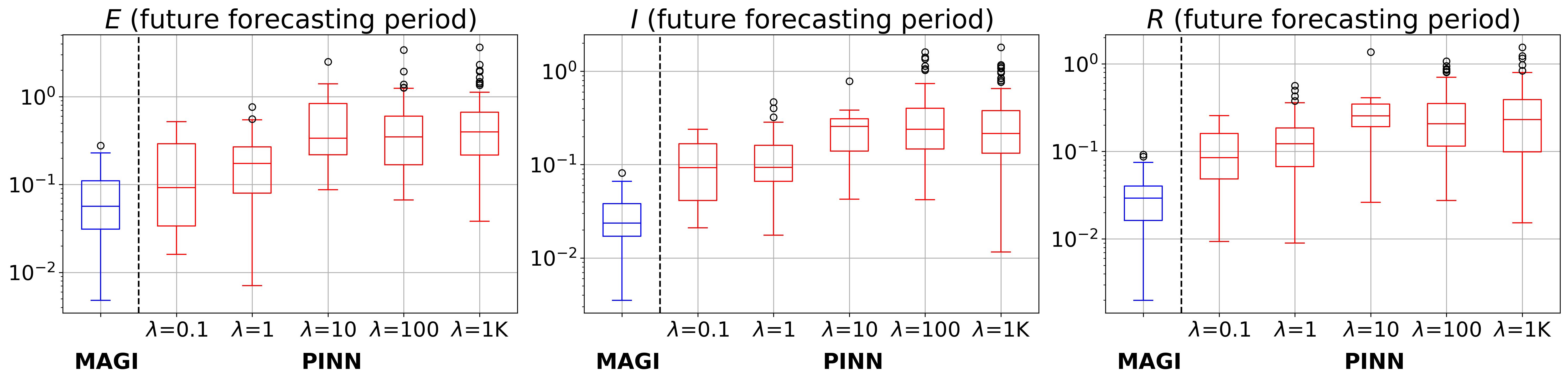}
        \caption{PINN results from the \texttt{DeepXDE} 
    implementation \citep{lu2021deepxde}, which 
    performs worse than the practitioner-style implementation 
    \citep{van2022physics} used in the main text for this SEIR 
    example. Boxplots showing the RMSE on the logarithm of the SEIR 
    system components across 100 datasets in the fully observed case. 
    Lower values indicate better performance. The y-axis is displayed 
    on a logarithmic scale for improved visualization. Top row: 
    in-sample trajectory reconstruction 
    (Equation~\eqref{eq:in-sample RMSE}); bottom row: future 
    forecasting (Equation~\eqref{eq:pred RMSE}). The three columns 
    correspond to the three system components $E$, $I$, and $R$. In 
    each panel, the leftmost boxplot corresponds to MAGI, and the 
    remaining boxplots correspond to best-case-per-$\lambda$ \texttt{DeepXDE} PINN under 
    different hyperparameter values~$\lambda$. The dashed vertical 
    line separates MAGI and PINN results.}
    \label{fig:si-full-seir-worse-pinn-boxplot_seir_traj_err_full_logscale}
\end{figure}

\clearpage

\begin{figure}[b!]
    \centering
    \includegraphics[width=1.0\linewidth]{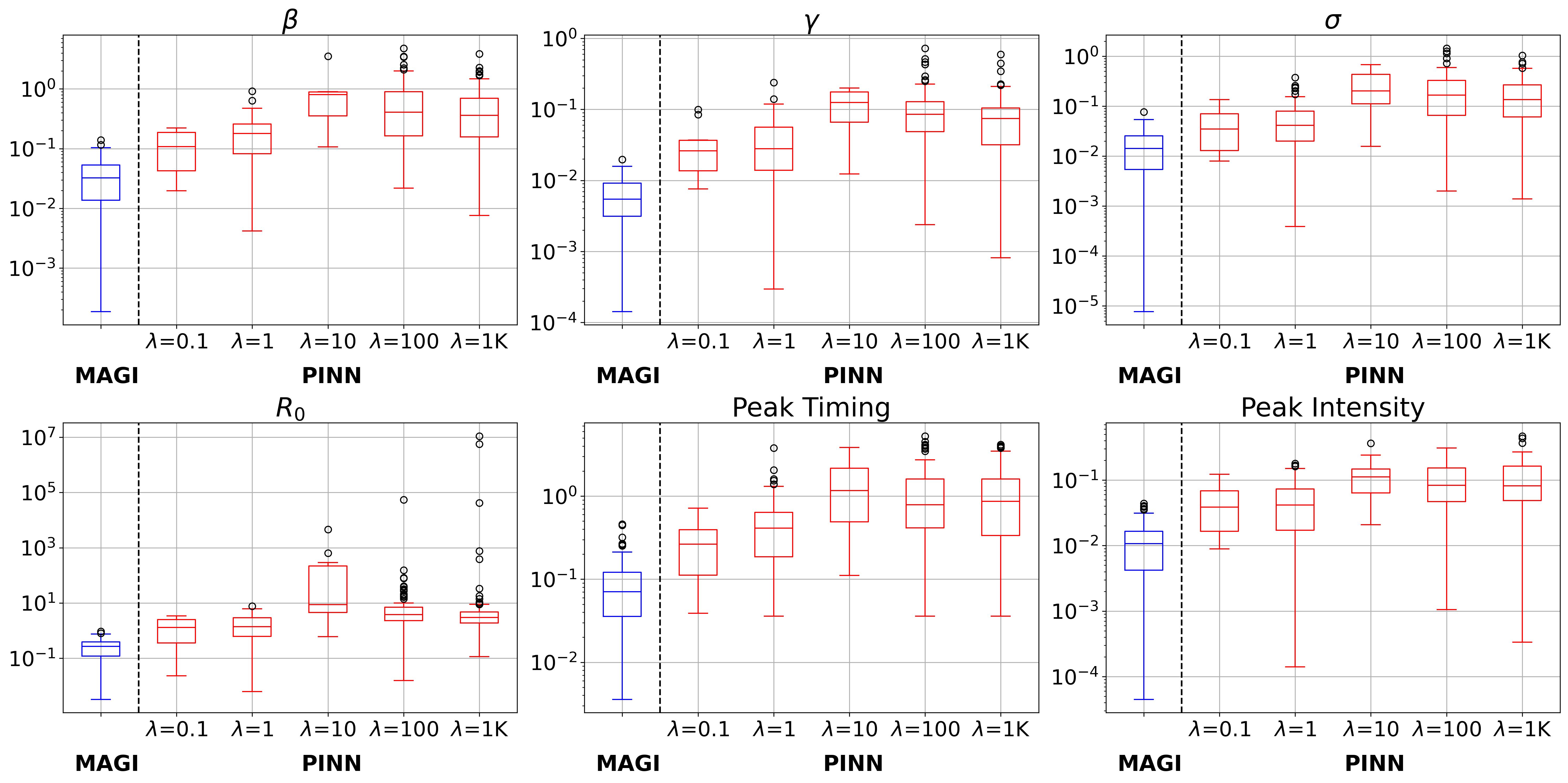}
    \caption{PINN results from the \texttt{DeepXDE} 
    implementation \citep{lu2021deepxde}, which 
    yields worse parameter estimation performance than the 
    practitioner-style implementation 
    \citep{van2022physics} used in the main text 
    for this SEIR example. Boxplots showing absolute parameter 
    estimation errors across 100 datasets in the fully observed case. 
    Lower values indicate better performance. The y-axis is displayed 
    on a logarithmic scale for improved visualization. Top row: errors 
    for the original SEIR parameters $\beta$, $\gamma$, and $\sigma$. 
    Bottom row: errors for $R_0$, peak timing, and peak 
    intensity---our three quantities of interest. In each panel, the 
    leftmost boxplot corresponds to MAGI, and the remaining boxplots 
    correspond to best-case-per-$\lambda$ \texttt{DeepXDE} PINN under different hyperparameter 
    values~$\lambda$. The dashed vertical line separates MAGI and PINN 
    results.}
    \label{fig:si-full-seir-worse-pinn-boxplot_param_err_full_logscale}
\end{figure}

\begin{figure}[H]
    \centering
    \includegraphics[width=1.0\linewidth]{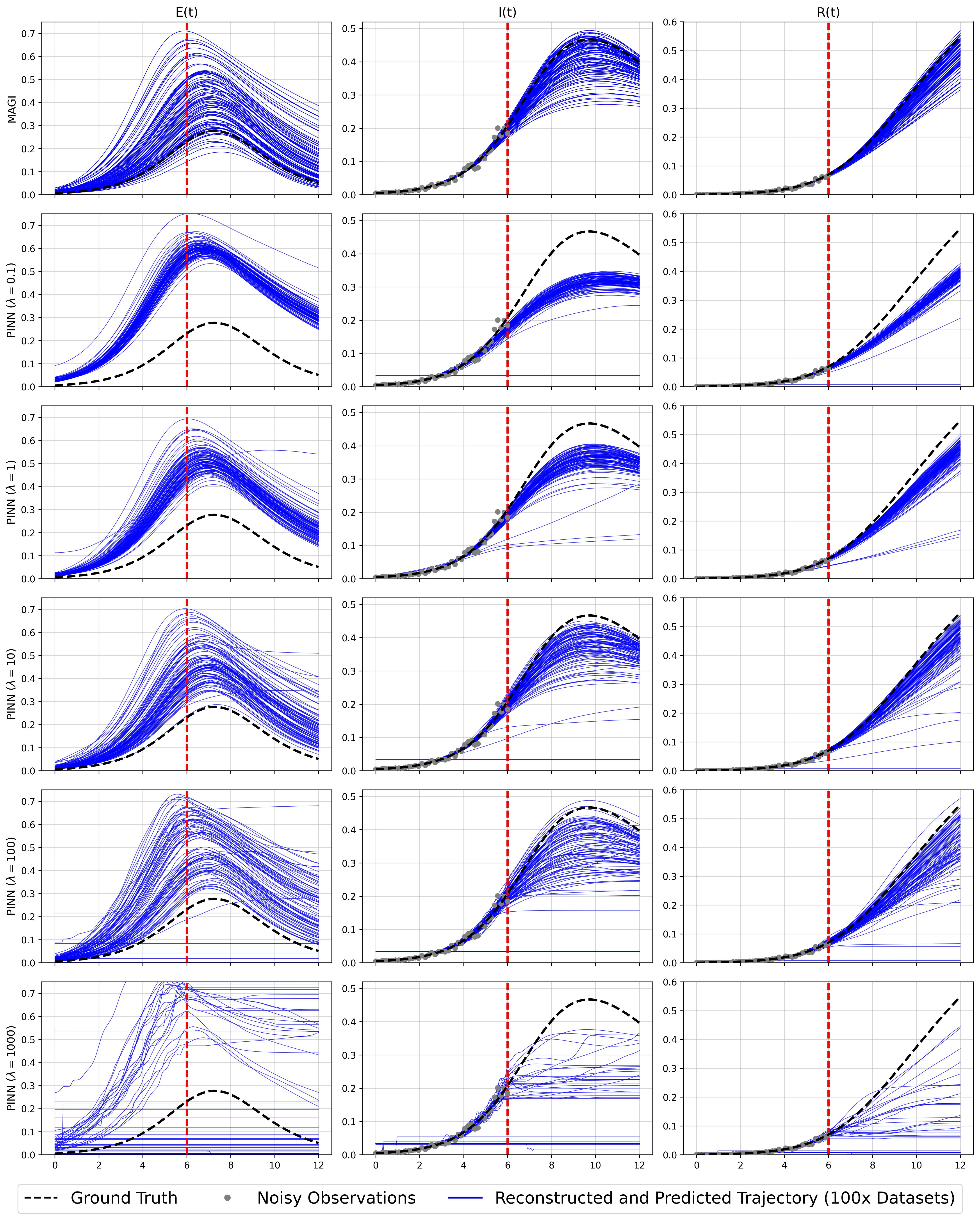}
    \caption{Trajectory reconstruction and prediction by best-case-per-$\lambda$ PINN and MAGI for the SEIR model in the missing component case. The dots show 
    one sample dataset (out of 100). The dashed black lines give the true curves, which are to be identified. The red dashed vertical line separates the in-sample observation period from the future forecasting period. Each solid blue curve is the estimate from one dataset. Top row: MAGI estimates. Lower five rows: best-case-per-$\lambda$ PINN estimates over all five tested $\lambda$ hyperparameter values.}
    \label{fig:seir_pinn_partial_plot}
\end{figure}

\clearpage
\begin{figure}[ht!]
    \centering
    \includegraphics[width=1.0\linewidth]{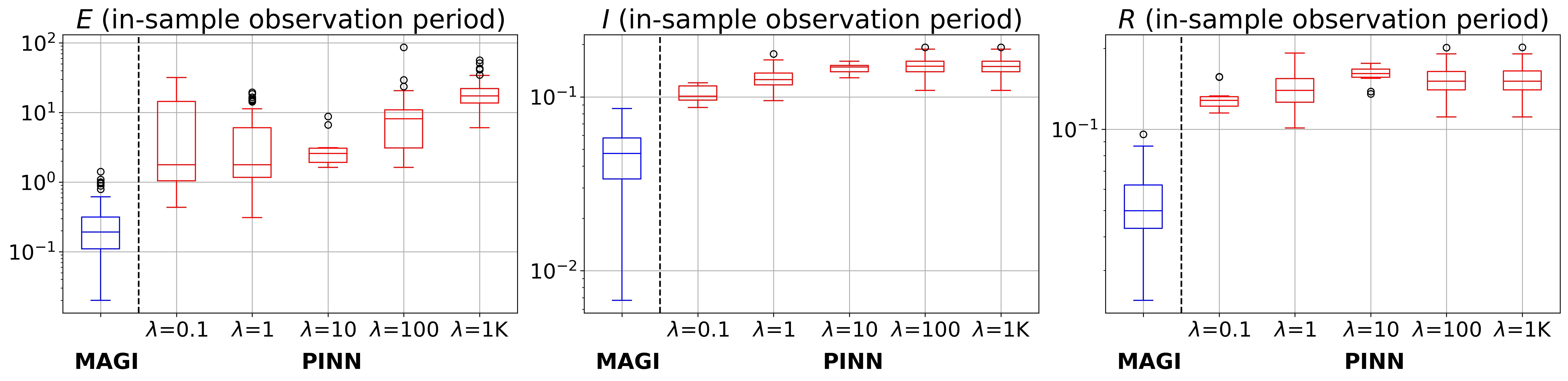}\\
    \includegraphics[width=1.0\linewidth]{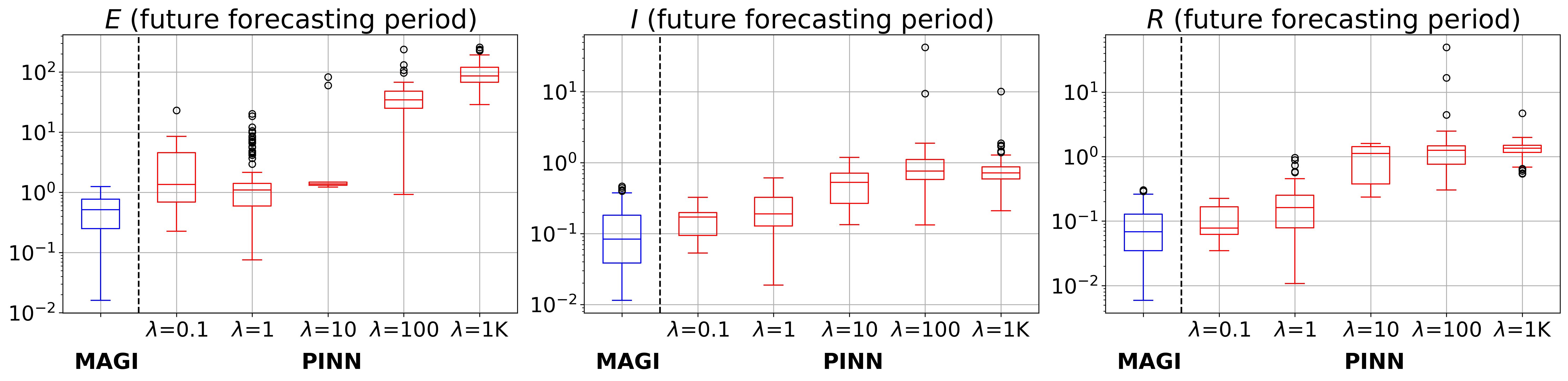}
    \caption{PINN results from the \texttt{DeepXDE} 
    implementation \citep{lu2021deepxde} for the 
    missing $E$ component case; as in the fully observed case, the 
    \texttt{DeepXDE}-based PINN performs worse than the 
    practitioner-style implementation used in the main text. Boxplots 
    showing the RMSE on the logarithm of the SEIR system components 
    across 100 datasets. The legend and layout are identical to 
    Figure~\ref{fig:boxplot_seir_traj_err_partial_logscale} in the 
    main text; see the caption there for details. In each panel, the 
    leftmost boxplot corresponds to MAGI, and the remaining boxplots 
    correspond to best-case-per-$\lambda$ \texttt{DeepXDE} PINN under different hyperparameter 
    values~$\lambda$. The dashed vertical line separates MAGI and PINN 
    results.}
    \label{fig:si-partial-seir-boxplot_seir_traj_err_partial_logscale}
\end{figure}
\clearpage
\begin{figure}[ht!]
    \centering
    \includegraphics[width=1.0\linewidth]{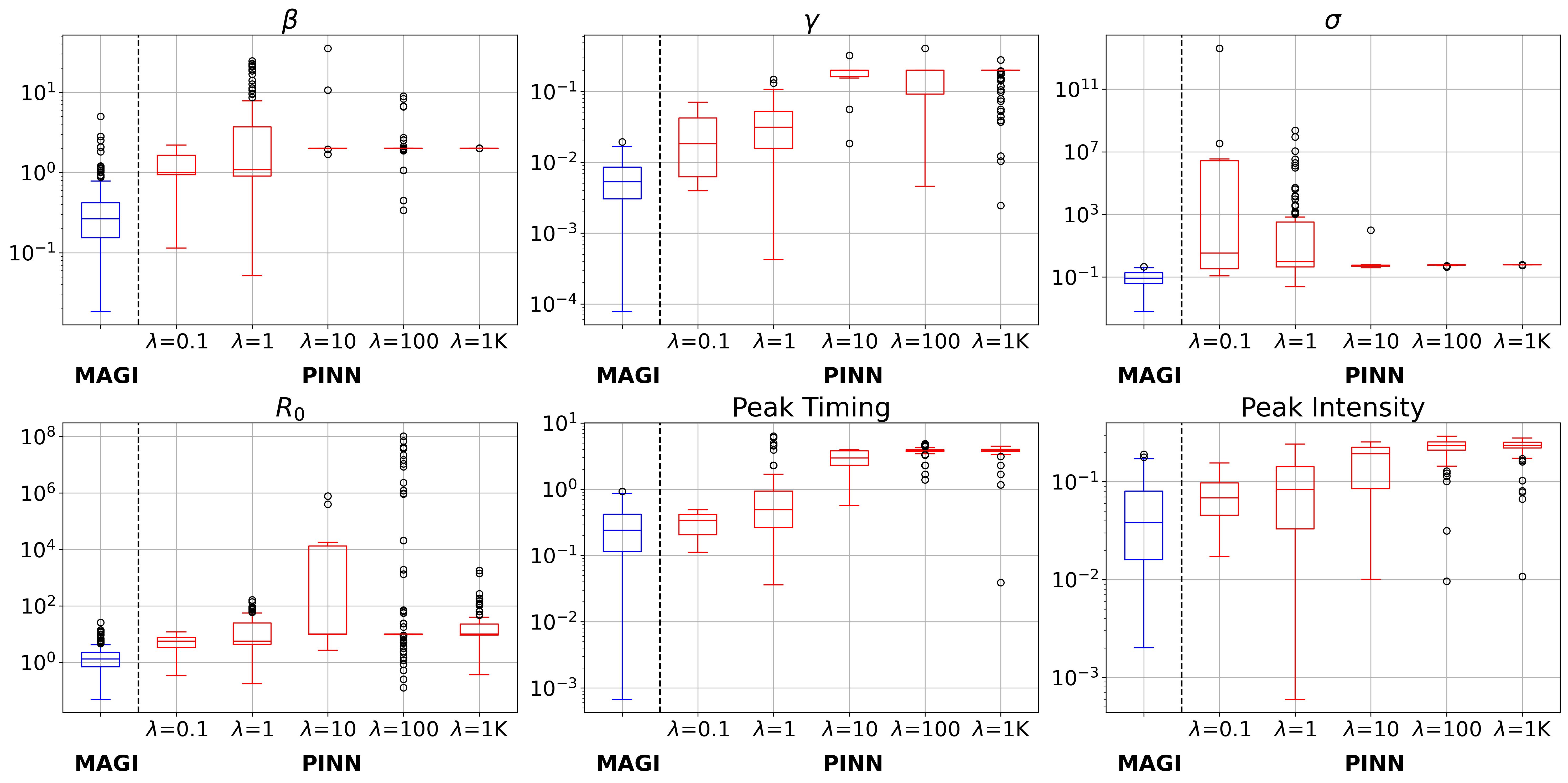}
    \caption{PINN results from the \texttt{DeepXDE} 
    implementation \citep{lu2021deepxde} for 
    parameter estimation in the missing $E$ component case; consistent 
    with the trajectory results, the \texttt{DeepXDE}-based PINN 
    yields worse performance than the practitioner-style 
    implementation used in the main text. Boxplots showing absolute 
    parameter estimation errors across 100 datasets. The legend and 
    layout are identical to 
    Figure~\ref{fig:boxplot_param_errors_partial} in the main text; 
    see the caption there for details. In each panel, the leftmost 
    boxplot corresponds to MAGI, and the remaining boxplots correspond 
    to best-case-per-$\lambda$ \texttt{DeepXDE} PINN under different hyperparameter 
    values~$\lambda$. The dashed vertical line separates MAGI and PINN 
    results.}
    \label{fig:si-partial-seir-boxplot_param_errors_partial}
\end{figure}

\section{Additional Lorenz chaotic (butterfly) results}
\begin{figure}[H]
    \centering
    \includegraphics[width=1.0\linewidth]{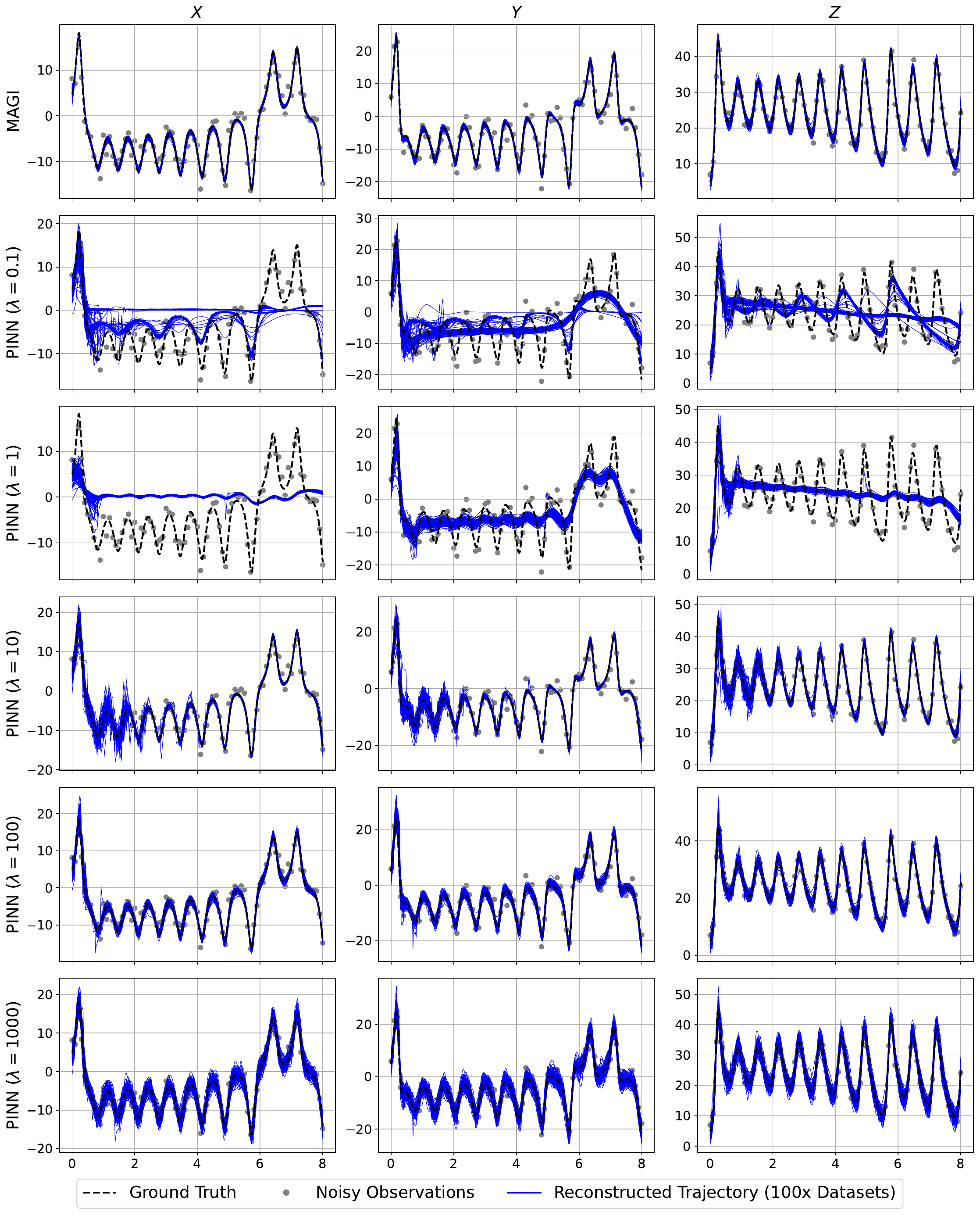} 
    \caption{Reconstructed trajectories by best-case-per-$\lambda$ PINN and MAGI on the Lorenz Chaotic (Butterfly) testbed across $100$ datasets, with one example dataset visualized. The dots show one sample dataset (out of 100). The dashed black lines give the true curves, which are to be identified. Each solid blue curve is the estimate from one data set. Top row: MAGI estimates. Lower five rows: best-case-per-$\lambda$ PINN estimates over all five tested $\lambda$ hyperparameter values.}
    \label{fig:chaotic_butterfly_traj_recons}
\end{figure}

\begin{figure}[H]
    \centering
    \includegraphics[width=1.0\linewidth]{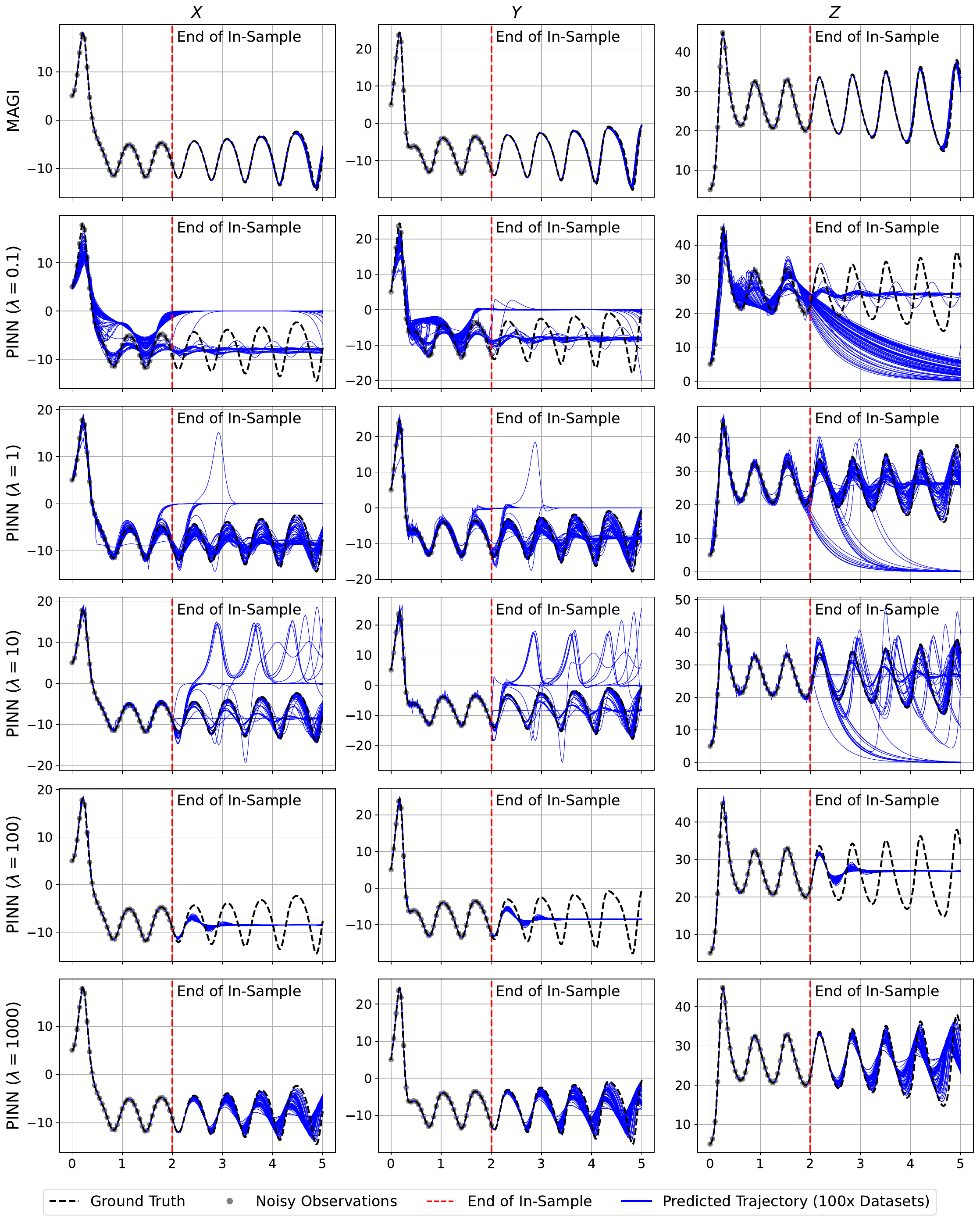} 
    \caption{Forecasted trajectories by best-case-per-$\lambda$ PINN and MAGI on the Lorenz Chaotic (Butterfly) testbed across $100$ datasets, with one example dataset visualized. The dots show one sample dataset (out of 100). The dashed black lines give the true curves, which are to be predicted. Each solid blue curve is the prediction from one data set. The red dashed vertical line separates the in-sample observation period from the future forecasting period. Top row: MAGI estimates. Lower five rows: best-case-per-$\lambda$ PINN estimates over all five tested $\lambda$ hyperparameter values.}
    \label{fig:chaotic_butterfly_forecasting}
\end{figure}

\clearpage
\begin{figure}[ht!]
    \centering
    \includegraphics[width=1.0\linewidth]{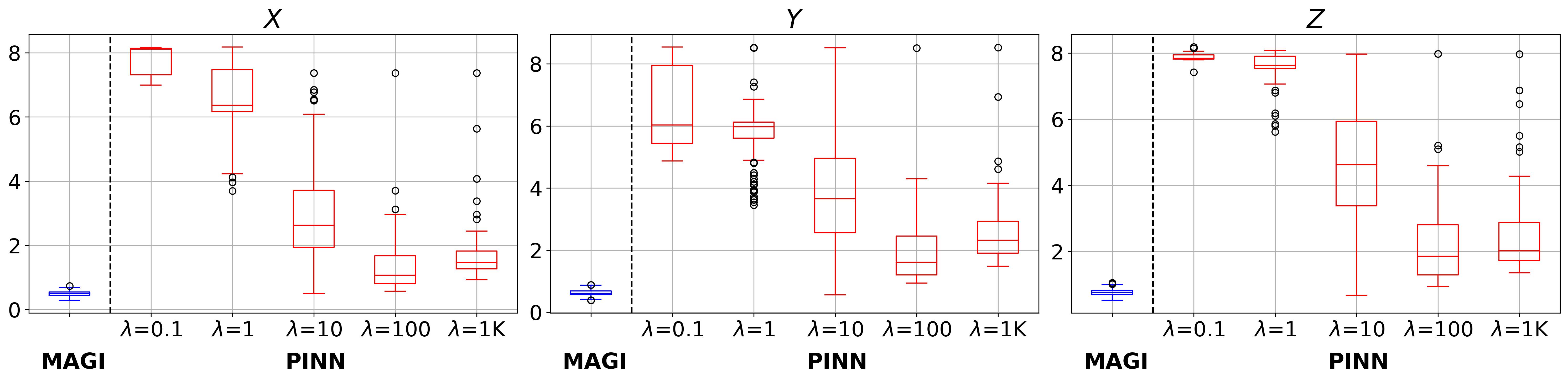} 
    \caption{PINN results from the practitioner-style 
    implementation \citep{van2022physics} for the 
    Lorenz Chaotic (Butterfly) testbed, which performs worse than 
    the \texttt{DeepXDE}-based PINN used in the main text. Boxplots 
    showing trajectory reconstruction RMSEs across $100$ datasets. 
    Lower values indicate better performance. The three panels 
    correspond to the system components $X$, $Y$, and $Z$. The legend 
    and panel layout are identical to 
    Figure~\ref{fig:chaotic_butterfly_traj_recons_metrics} in the main 
    text; see the caption there for details. In each panel, the 
    leftmost boxplot corresponds to MAGI, and the remaining boxplots 
    correspond to 
    practitioner-style PINN under different hyperparameter 
    values~$\lambda$. The dashed vertical line separates MAGI and PINN 
    results.}
    \label{fig:si-lorenz-chaotic_butterfly_traj_recons_metrics}
\end{figure}
\begin{figure}[ht!]
    \centering
    \includegraphics[width=1.0\linewidth]{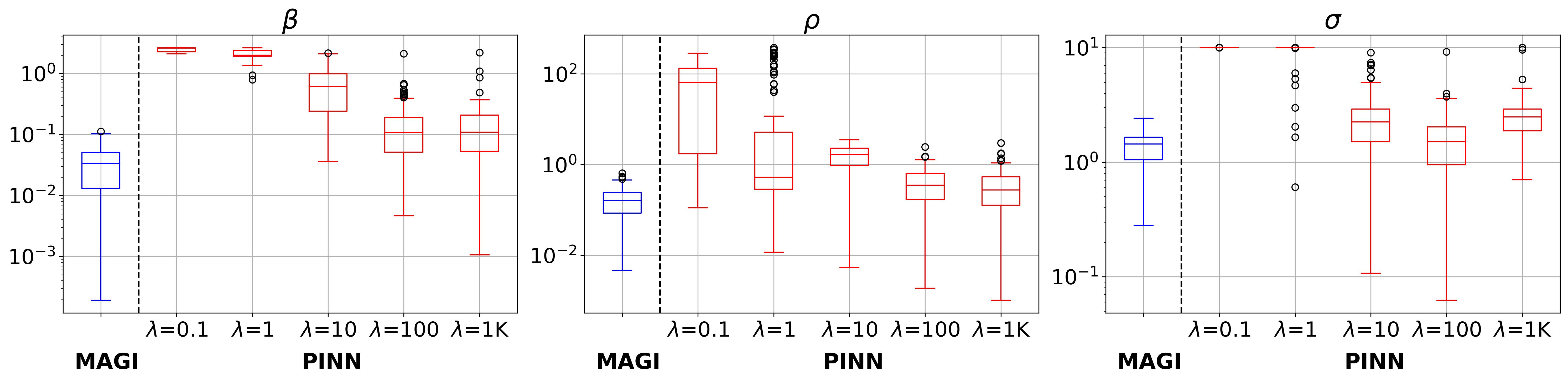} 
    \caption{PINN results from the practitioner-style 
    implementation \citep{van2022physics} for 
    parameter estimation on the Lorenz Chaotic (Butterfly) testbed, 
    which yields worse performance than the \texttt{DeepXDE}-based 
    PINN used in the main text. Boxplots showing parameter inference 
    absolute errors across $100$ datasets. Lower values indicate 
    better performance. The three panels correspond to the system 
    parameters $\beta$, $\rho$, and $\sigma$. The legend and panel 
    layout are identical to 
    Figure~\ref{fig:chaotic_butterfly_param_inf_metrics} in the main 
    text; see the caption there for details. In each panel, the 
    leftmost boxplot corresponds to MAGI, and the remaining boxplots 
    correspond to
    practitioner-style PINN under different hyperparameter 
    values~$\lambda$. The dashed vertical line separates MAGI and PINN 
    results.}
    \label{fig:si-lorenz-chaotic_butterfly_param_inf_metrics}
\end{figure}

\newpage

\section{Results from stable (transient chaos) regime}
\label{subsec:lorenz_stable_transient_chaos_results}

\begin{figure}[ht!]
    \centering
    \includegraphics[width=1.0\linewidth]{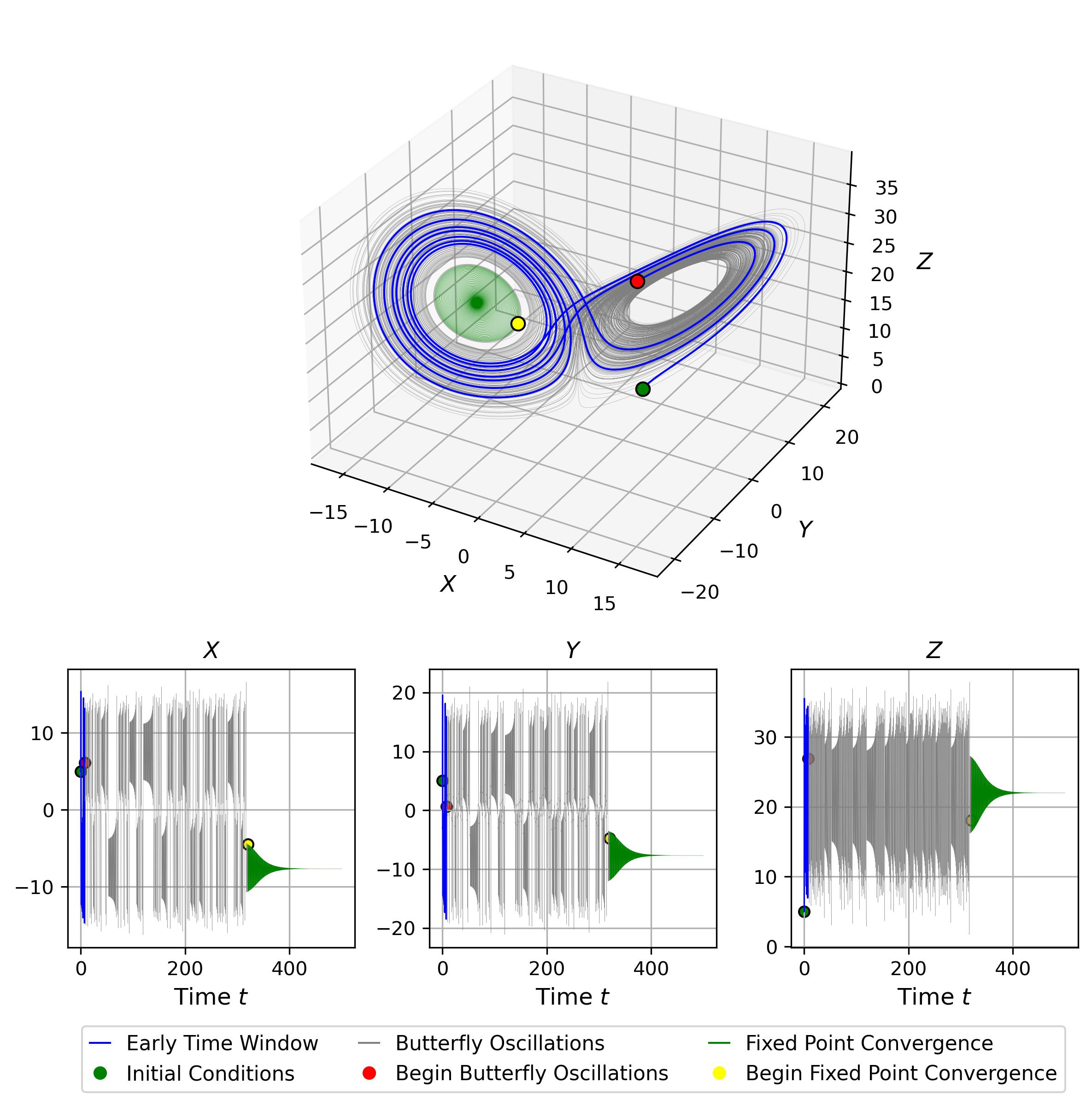} 
    \caption{The Stable (Transient Chaos) regime of the Lorenz system with parameters $\bm{\theta} = (\beta, \rho, \sigma) = (\frac{8}{3}, 23, 10)$ and initial condition $(X(0), Y(0), Z(0)) = (5, 5, 5)$.}
    \label{fig:stable_transient_chaos}
\end{figure}

In the Lorenz system, some parameter settings can lead to trajectories that initially exhibit chaotic butterfly-shaped behavior, before eventually converging to a stable fixed point (i.e., ``preturbulence'' or ``transient chaos''), or lead to trajectories that perpetually alternate between chaotic and stable-looking behavior.

Our second testbed, depicted in SI Figure \ref{fig:stable_transient_chaos}, which we refer to as ``Stable (Transient Chaos),'' initially shows butterfly oscillatory behavior, before eventually converging to a fixed point. %
This regime is mathematically stable \citep{sparrow2012lorenz}, with parameters $\bm{\theta} = (\beta, \rho, \sigma) = (\frac{8}{3}, 23, 10)$ and initial condition $(X(0), Y(0), Z(0)) = (5, 5, 5)$.

Mirroring the ``Chaotic (Butterfly)'' testbed, for ``Stable (Transient Chaos),'' we generate 100 independent data sets. The observation window is from $t \in [0, 8]$, and a total of 81 equally-spaced observations are generated within this interval for each data set, which gives $I_{\text{obs}}$, with $5\%$ additive Gaussian noise injected. %
For both PINN and MAGI, we use a shared discretization set of $I$ as $321$ evenly-spaced time steps in $t \in [0, 8]$ to evaluate the physics-based loss component (see Equation \eqref{eq:pinn}) and $W_I$ (see Equation \eqref{eq:WI}, respectively.  All PINN and MAGI implementation details are identical to those described in Section \ref{sec:lorenz-experimental-setup}.

In the Stable (Transient Chaos) regime, as shown in SI Figure \ref{fig:stable_transient_chaos}, MAGI's inferred trajectories continue to closely hug the ground-truth trajectory, with low trial-to-trial variability. In contrast, even the best-performing PINNs' inferred trajectories appear noticeably ``fuzzier,'' with subtantially larger variability over trials. We also observe flat-line trajectory collapse for PINNs with smaller $\lambda$ values. In general, MAGI and PINN behaviors are very similar to those in in the Chaotic (Butterfly) regime.

\begin{figure}[ht!]
    \centering
    \includegraphics[width=1.0\linewidth]{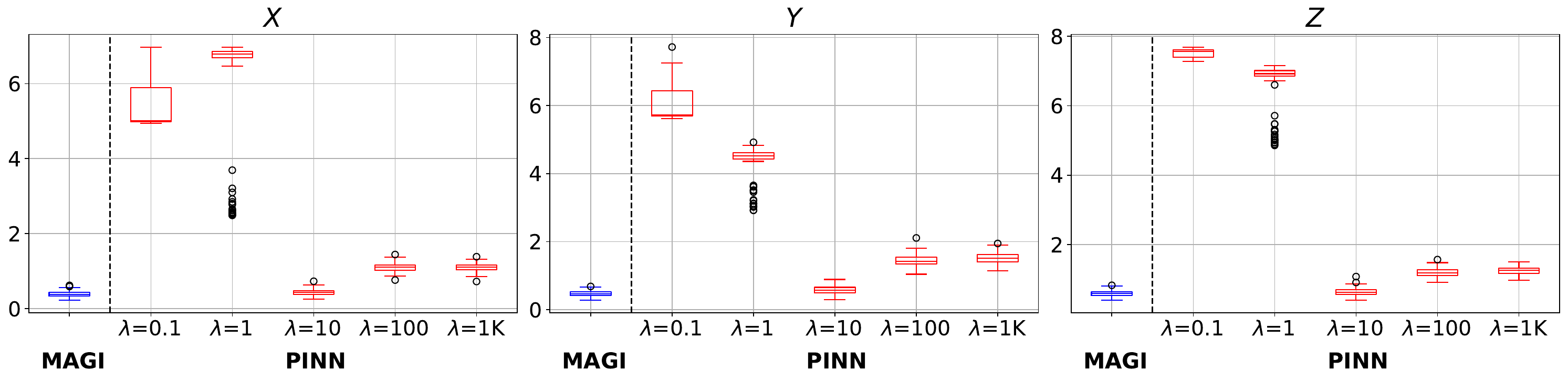} 
    \caption{Boxplots showing trajectory reconstruction RMSEs across $100$ datasets on the Lorenz Stable (Transient Chaos) testbed. Lower values indicate better performance. The three panels correspond to the three system components $X, Y$, and $Z$. In each panel, the leftmost blue boxplot is for MAGI, while the remaining red boxplots are for PINN (best case per $\lambda$) under different $\lambda$ hyperparameter settings. The dashed vertical line separates the MAGI and PINN results.}
    \label{fig:stable_transient_chaos_traj_recons_metrics}
\end{figure}

A closer analysis of trajectory RMSE, presented in the boxplots of SI Figure \ref{fig:stable_transient_chaos_traj_recons_metrics}, confirms this visual impression. MAGI achieves the smallest error, while PINN at \(\lambda = 10\) shows an error closer to MAGI. The physics-emphasizing PINN variants with small \(\lambda\) continue to produce trivial flat-line predictions, which significantly contribute to the large errors observed in the boxplots.

\begin{figure}[ht!]
    \centering
    \includegraphics[width=1.0\linewidth]{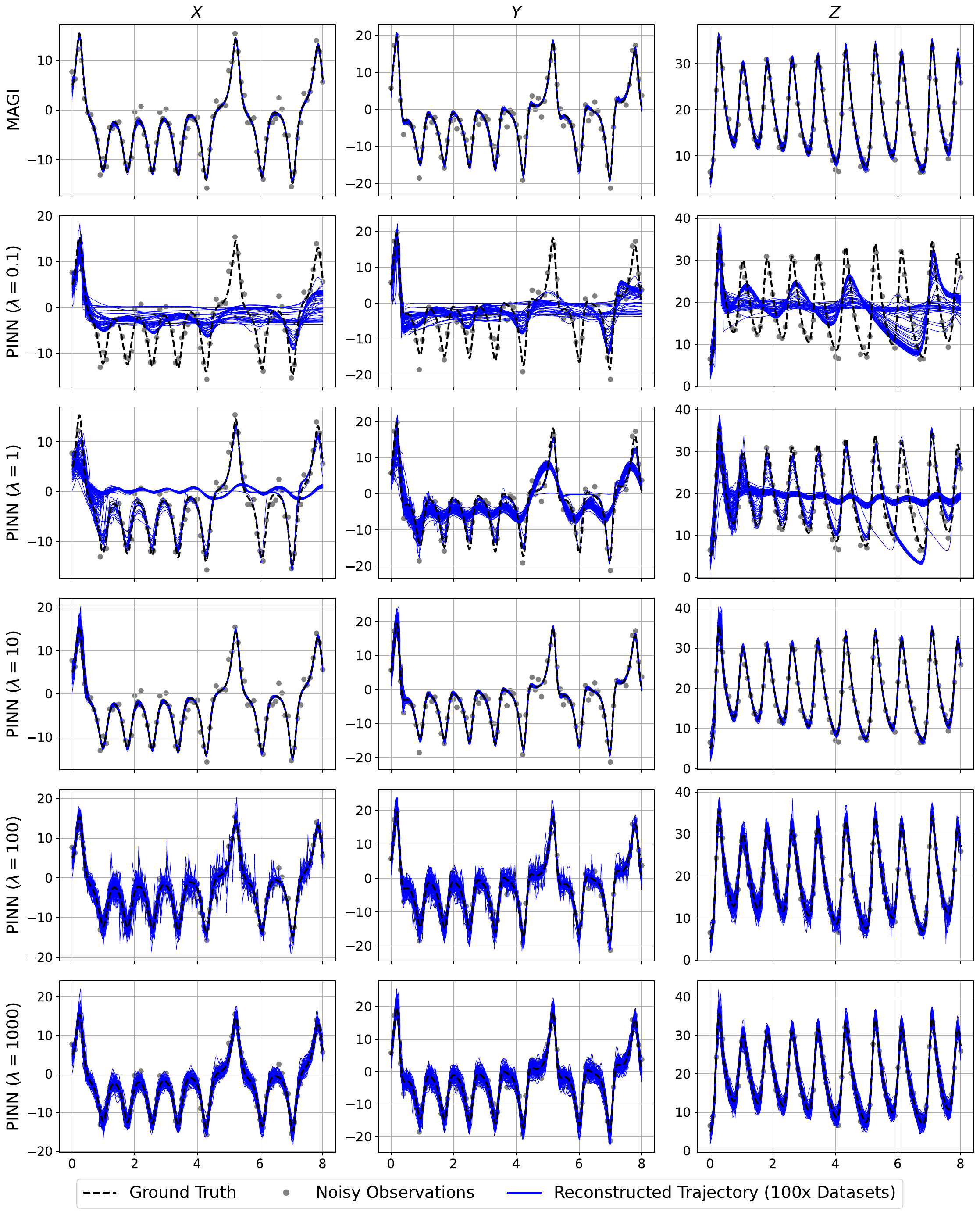} 
    \caption{MAGI and PINN (best case per $\lambda$) reconstructed trajectories on the Lorenz Stable (Transient Chaos) testbed across $100$ datasets, with one example dataset visualized. The dashed black lines give the true curves, which are to be inferred. Each solid blue curve is the estimate from one data set. Top row: the $100$ MAGI posterior means for each dataset. Lower five rows: best-case-per-$\lambda$ PINN estimates over all five tested $\lambda$ hyperparameter values. We note that the 100 reconstructed trajectories for MAGI are all indeed in the subplots: they are all on top of each other, indicating very low variance across datasets.}
    \label{fig:stable_transient_chaos_traj_recons}
\end{figure}

\begin{figure}[ht!]
    \centering
    \includegraphics[width=1.0\linewidth]{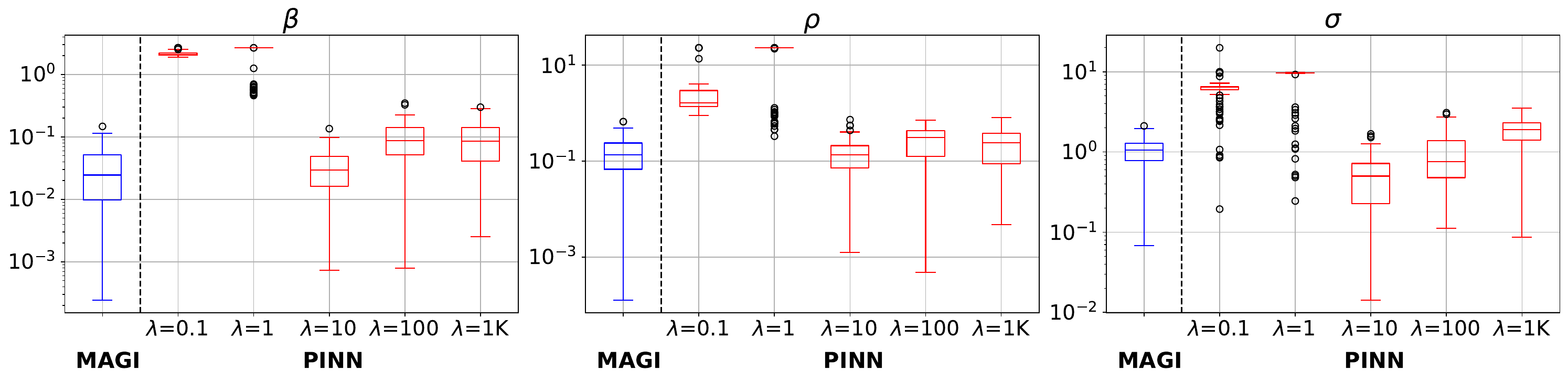} 
    \caption{Boxplots showing parameter inference absolute errors across $100$ datasets on the Lorenz Stable (Transient Chaos) testbed. Lower values indicate better performance. The three panels correspond to the three system parameters $\beta, \rho$, and $\sigma$. In each panel, the leftmost blue boxplot is for MAGI, while the remaining red boxplots are for (best case per $\lambda$) under different $\lambda$ hyperparameter settings. The dashed vertical line separates the MAGI and PINN results.}
    \label{fig:stable_transient_chaos_param_inf_metrics}
\end{figure}

For parameter inference, SI Figure \ref{fig:stable_transient_chaos_param_inf_metrics} presents the boxplots of absolute errors for each model parameter. Results are very similar to those of the Chaotic (Butterfly) setting: MAGI outperforms or performs comparably to the strongest PINN variants on $\beta$ and $\rho$, but underperforms slightly on $\sigma$.

\begin{figure}[ht!]
    \centering
    \includegraphics[width=1.0\linewidth]{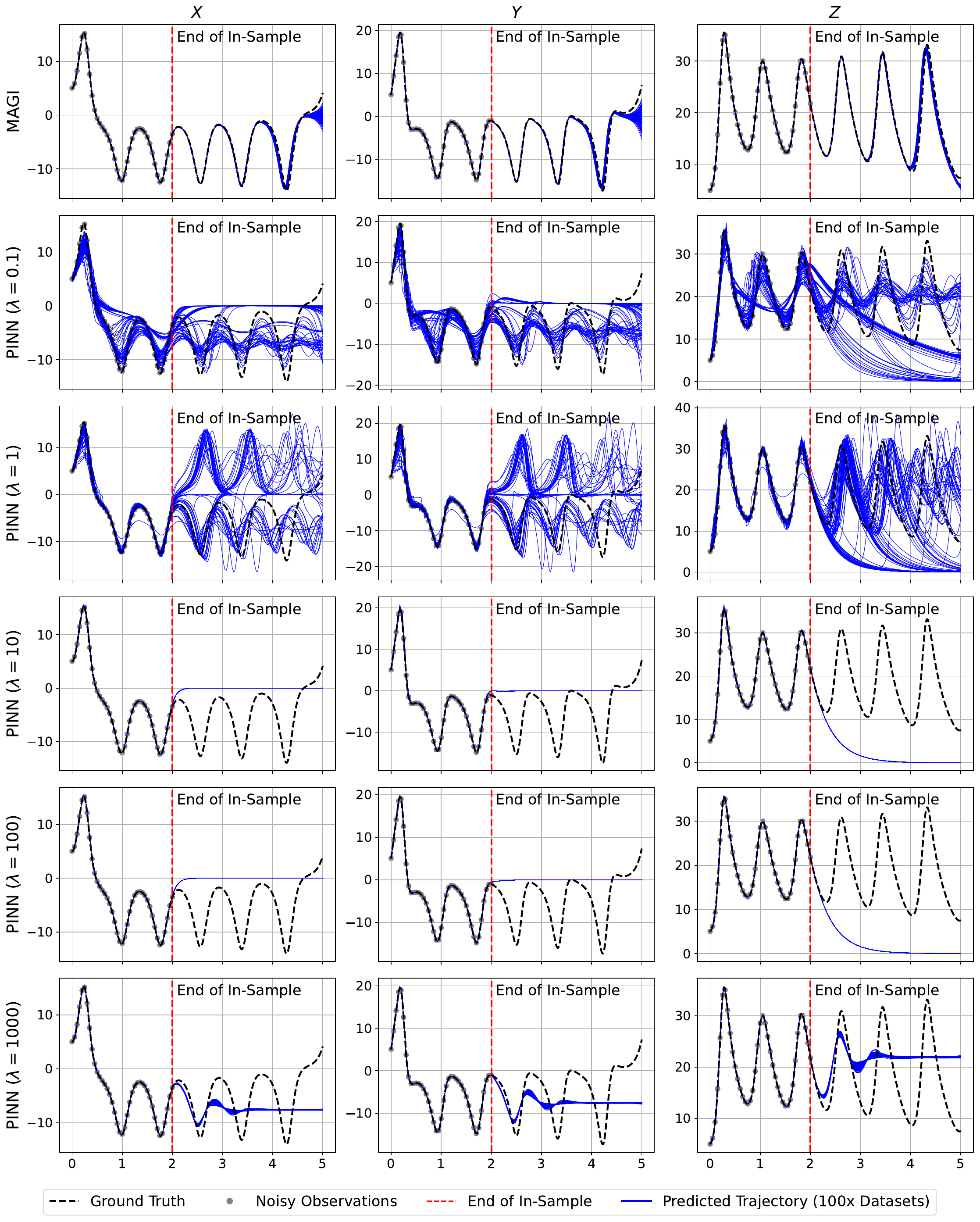} 
    \caption{MAGI and PINN (best case per $\lambda$) forecasted trajectories on the Stable (Transient Chaos) testbed across $100$ datasets, with one example dataset visualized. The dashed black lines give the true curves, which are to be predicted. Each solid blue curve is the prediction from one data set. The red dashed vertical line separates the in-sample observation period from the future forecasting period. Top row: the $100$ MAGI posterior means for each dataset. Lower five rows: best-case-per-$\lambda$ PINN estimates over all five tested $\lambda$ hyperparameter values.}
    \label{fig:stable_transient_chaos_forecasting}
\end{figure}

The future prediction results in SI Figure \ref{fig:stable_transient_chaos_forecasting} are also similar to those of the Chaotic (Butterfly) regime. Here, all PINN variants exhibit a tendency to collapse towards flat-line predictions as the forecast extends further into the future. In contrast, MAGI demonstrates consistent and accurate forecasting over the entire interval from \(t = 2\) to \(t = 5\). However, the forecasts begin to diverge noticeably near the end of the interval at \(t = 5\), reaching the limit of this Lorenz system example. Comparing the results between the two testbeds, one may hypothesize that forecasting is more tractable on the Chaotic (Butterfly) regime than the Stable (Transient Chaos) regime, the former which we analyzed at the end of Section \ref{sec: Lorenz}. 

Overall, from the above results, it is clear that the Chaotic (Butterfly) and Stable (Transient Chaos) testbeds tell very similar stories about the performance differences between MAGI and PINN.

\end{document}